\documentclass{article}
\usepackage{fullpage}
\usepackage{thm-restate}
\usepackage{algorithm,algorithmic}
\usepackage{adjustbox}
\usepackage[round]{natbib}
\usepackage{amsmath,amssymb,bm,enumerate,mathrsfs,mathtools}
\usepackage{latexsym,color,verbatim,multirow}
\usepackage{graphicx}
\usepackage{caption}
\usepackage{subcaption}
\usepackage{enumitem}

\newtheorem{theorem}{Theorem}[section]

\newtheorem{assumptions}[theorem]{Assumption}

\newtheorem{lemma}[theorem]{Lemma}

\newtheorem{example}[theorem]{Example}

\def\bB {{\bf B}}
\def\bX {{\bf X}}
\def\bM {{\bf M}}
\def\bV {{\bf V}}

\def\bD {{\bf D}}
\def\bZ {{\bf Z}}
\def\bU {{\bf U}}
\def\bW {{\bf W}}
\def\bS {{\bf S}}
\def\bP {{\bf P}}
\def\bA {{\bf A}}
\def\bE {{\bf E}}
\def\bDelta {{\bf \Delta}}
\def\bSigma {{\bf \Sigma}}
\def\bGamma {{\bf \Gamma}}
\newcommand{\real}{\mathbb{R}}

\newcommand{\Var}{\operatorname{Var}}
\DeclareMathOperator*{\argmin}{arg\,min}
\newcommand{\Id}{\mathbf{Id}}

\begin{document}

\title{Regularization for supervised learning via the ``hubNet'' procedure}
\author{Leying  Guan \thanks{Dept. of  Statistics,
    Stanford Univ, leying.guan@gmail.com},  Zhou Fan\thanks{Dept. of  Statistics,
    Stanford Univ, zhoufan@stanford.edu}, and  Robert Tibshirani \thanks{Depts. of Biomedical Data Sciences, and Statistics,
    Stanford Univ, tibs@stanford.edu}\\
Stanford University } 

\maketitle

\begin{abstract}
We propose a new method for supervised learning. The {\em hubNet }procedure fits a hub-based graphical model
to the predictors,  to estimate the amount of  ``connection'' that each predictor has with other predictors.
This yields a set of predictor weights that are then used in a regularized regression such as the lasso or elastic net.
The resulting procedure is easy to implement, can sometimes yields higher prediction accuracy that the lasso,
and can give insights into the underlying structure of the predictors.
HubNet can  also be generalized seamlessly to other supervised problems such as
regularized logistic regression (and other GLMs), Cox's proportional hazards
model, and nonlinear procedures such as random forests and boosting.
We prove some recovery results under a specialized model and illustrate the  method on real and simulated data.

\end{abstract}

\section{Introduction}

We consider the usual linear regression model: Given $n$ realizations of $p$
predictors $\bX=\{x_{ij}\}$ for $i=1,2,\ldots,n$ and $j=1,2,\ldots,p$, the
response $Y=(y_1,\ldots,y_n)$ is modeled as
\begin{equation}
y_i=\beta_0 + \sum_j x_{ij} \beta_j +\epsilon_i
\end{equation}
with $\epsilon \sim (0,\sigma^2)$. The
ordinary least squares (OLS) estimates of $\beta_j$ are obtained by minimizing the residual sum of squares.
There has been much work on regularized estimators that offer an advantage over the OLS estimates,
both in terms of accuracy of prediction on future data and interpretation of the fitted model.
One major focus has been on the {\em lasso} \citep{Ti96}, which minimizes
\begin{equation}
J(\beta_0,\beta)=\frac{1}{2}\|Y-\beta_0-\bX \beta\|_2^2+\lambda \|\beta\|_1
\end{equation}
where $\beta=(\beta_1, \ldots,\beta_p)$, and the tuning parameter  
$\lambda \geq 0$ controls the sparsity of the final model. This parameter is often selected by cross-validation.
The objective function $J(\beta_0,\beta)$ is convex, which means that the solutions can be found efficiently even for very large
$n$ and $p$, in contrast to combinatorial methods like best subset selection.
A body of mathematical work shows that under certain conditions, the lasso often will provide good recovery of the underlying true model
and will produce predictions that are mean-square consistent
\citep{KF2000,meinshausen2006high,zhao2006model,bunea2007sparsity,
zhang2008sparsity,meinshausen2009lasso,bickel2009simultaneous,
wainwright2009sharp}.
The {\em elastic net}  of  \citet{enet} generalizes the lasso by adding an
$\ell_2$ penalty,
\begin{equation}
\frac{1}{2}\|Y-\beta_0-\bX\beta\|_2^2
+\lambda (\alpha \|\beta\|_1 +(1-\alpha) \|\beta\|_2^2),
\end{equation}
where $\alpha \in [0,1]$ is a second tuning parameter. This approach sometimes
yields lower prediction error than the lasso, especially in settings with
highly correlated predictors.

\citet{zou2006a} introduced the {\em adaptive lasso}, which minimizes
\begin{equation}\label{eq:adaptivelasso}
\frac{1}{2}\|Y-\beta_0-\bX\beta\|_2^2+\lambda \sum_j  w_j|\beta_j|
\end{equation}
for feature weights $w_j$. The feature weights can be chosen in various ways: 
For example, when $n>p$, we can first compute the OLS estimates $\hat\beta_j$ 
and then set $w_j=1/|\hat\beta_j|$. For $p>n$, we can set $w_j$ by first
computing univariate regression coefficients \citep{huang2008adaptive}.
Other similar ``two-step'' procedures include variants of the non-negative
garrote \citep{breiman1995better,yuan2007non} and the adaptive elastic net
\citep{zou2009adaptive}. We have found that one
less than ideal property  of the adaptive lasso is that there seems to be no underlying generative model that leads to its feature weighting. Perhaps as a result, it is difficult even to simulate a dataset 
 that shows substantial gains for the method, relative to the usual lasso.

In this paper, we provide a new perspective by choosing weights in the adaptive
lasso in an unsupervised manner. All of the above two-step procedures select
weights by computing an initial estimate $\hat\beta$ using the response $Y$.
We instead propose
to use the partial correlations of the features in $\bX$ to select good weights.
We postulate a conceptual model in which there is a core subset $S$ of ``hub''
features that explains both the other features and $Y$.
For example, each member of $S$  might be the RNA or protein expression of a
``driver'' gene in a pathway which simultaneously influences other gene
expressions and the phenotype under study.
Our method, called {\em hubNet}, fits an (unsupervised) graphical model to the
features in a way that tries to discover these ``hubs''.
These features are then given higher weight in the adaptive lasso.
The hubNet procedure can sometimes  yield lower prediction error and better
support recovery than the lasso, and the discovered hubs can provide
insight on the underlying structure of the data.

 This paper is organized as follows. In Section \ref{sec:hubNet}  we introduce our underlying model and the hubNet procedure.
 Simulation studies are presented in Section \ref{sec:sim}, while Section
\ref{sec:real} examines applications to real datasets.
 Some theoretical results on the recovery of the underlying model are given in
Section \ref{sec:theory}. Further topics are discussed
 in Section \ref{sec:further}, such as extensions to random forests
and post-selection inference. Section \ref{sec:hubmethodcomparison} compares our method of identifying hubs
with an alternative approach.

\subsection {Illustrative example: Olive oil data}

The data for this example, from \citet{forina1983}, consists of measurements of   8 fatty acid concentrations for 572 olive oils,  with each olive oil classified into one of two geographic regions. The goal is to determine the geographic region based on these 8 predictors. We randomly divided the data into training and test sets of equal size. The predictors are:

\begin{enumerate}[itemsep=-1mm]
\item Palmitic Acid 
\item  Palmitoleic Acid 
\item  Stearic Acid 
\item  Oleic Acid 
\item  Linoleic Acid 
\item Linolenic Acid 
\item Arachidic Acid 
\item Eicosenoic Acid
\end{enumerate}

\begin{figure}
\begin{center}
\includegraphics[width=\textwidth]{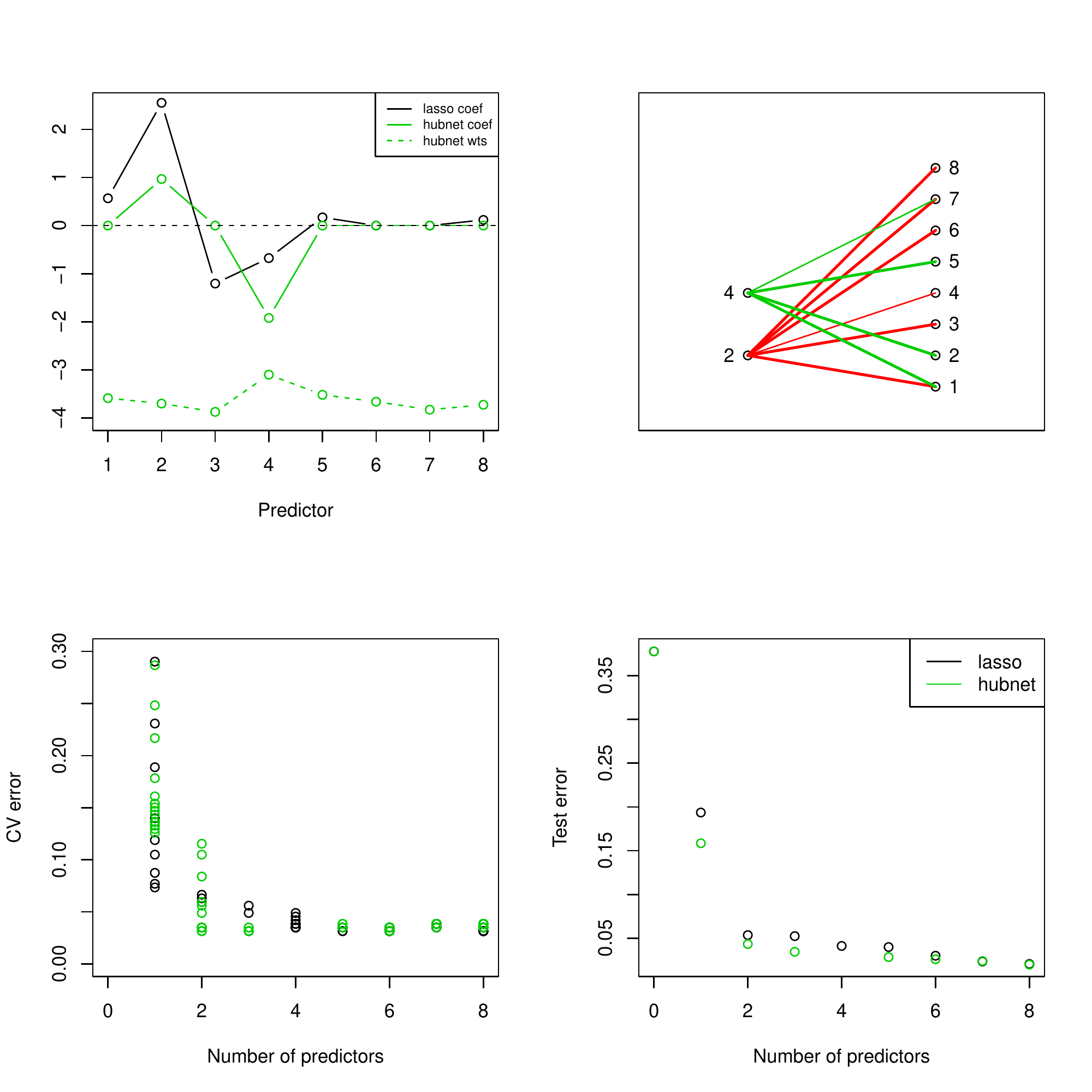}
\caption[fig:olive]{\em Results for olive oil data. Top left panel shows coefficients from lasso (black),
hub weights (broken green line) and resulting coefficients from hubNet (solid green).   hubNet chooses
predictors 2 (palmitoleic acid) and 4 (oleic acid), having connections to other predictors as depicted  in the  top right panel. The boldness of the link
corresponds to the strength of the association.  The bottom panels show the cross-validation and test error for the lasso and hubNet.}
\label{fig:olive}
\end{center}
\end{figure}

Results from hubNet and lasso-regularized logistic regression are given in Figure \ref{fig:olive} with details in the caption. (Extension of hubNet to logistic regression is straightforward and discussed in Section \ref{sec:glm}.) HubNet focuses on just two predictors---2 and 4, which have apparent connections
to the other six. In the process, it yields a more parsimonious model than the
lasso, with perhaps a lower CV and test error.

\section{The hubNet procedure}

\label{sec:hubNet}
Let $Y=(y_1, \ldots,y_n)$ and let $\bX=\{x_{ij} \}$ be the $n\times p$ matrix
of features.
Define the core set $S$ to be a subset of $\{1,2,\ldots p\}$, with corresponding
feature matrix $\bX_S$. Our proposal is based on the following model:
\begin{eqnarray}
Y &= &\beta_0+\bX_S\beta + \epsilon \label{eqn:linearmodel}\\
X_j &=& \bX_S\Gamma_j + \epsilon_j,\;\;j \notin S \label{eqn:model}
\end{eqnarray}
where  each $\Gamma_j$ is an $s\times 1$ coefficient vector.
This model postulates  that the outcome $Y$ is a function of an (unknown) core set of predictors $S$, and that the predictors not in $S$ are also
a function of this same core set.

If this model holds, even approximately, then we can examine the partial
correlations among the features to determine the features more likely to belong to this core set $S$,
and hence do a better job of predicting $Y$.  
Following this logic, our proposal for estimating $\beta$ in (\ref{eqn:linearmodel})  consists of three steps:
\bigskip

\begin{algorithm}
\medskip

\centerline{\bf The hubNet procedure}

\begin{enumerate}
\item Fit a model of the form $\bX\approx \bX\bB$ with $\bB_{ii}=0$ using the
 ``edge-out'' procedure detailed in Section \ref{sec:edgeout} below.
Note that $\Gamma_j$ in the generating model (\ref{eqn:model}) correspond to  coefficients of $\bB$
in rows $S$ and columns $S^C$.

\item Let $s_j=\sum_j |\hat\bB_{ij}|, j=1,2,\ldots,p$, and
construct feature weights 
\begin{equation}
w_j=1/s_j\,.
\label{eqn:wtfamily}
\end{equation}
\item Fit the adaptive lasso using predictors and feature
weights $w_j$ (e.g., using $w_j$ as ``penalty factors'' in the {\tt glmnet} R
package.) [If $s_j=0$, then $w_j=\infty$ and $X_j$ is not used.]
\end{enumerate}

\medskip

\end{algorithm}

\bigskip

The hubNet procedure has a number of attractive features:

\begin{description}
\item{(a)} The construction of weights is completely unsupervised,  separating it from the fitting of the response model in step 3. Thus for example, cross-validation can be applied in step 3 and we can use cross-validation to 
choose between hubNet and lasso for a given problem. In addition, 
tools for post-selection inference for the lasso can be directly applied.
\item{(b)} The supervised fitting in step 3 is simply a lasso (or elastic net)
with feature weights, hence fast off-the-shelf solvers can be used.
\item{(c)} Examination of the estimated hub structure for the chosen predictors can shed light on the structure of the final model.
\item{(d)} The procedure can be directly applied to generalized regression
settings, such as generalized linear models and the proportional hazards model
for survival data, using an appropriate method in step 3.
\end{description}

The challenging task of the hubNet procedure is step 1.
For this, one might use the graphical lasso, which produces a sparse estimate of
the inverse covariance matrix, corresponding to an edge-sparse  feature graph.
But we would like an estimate that encourages the appearance of hub nodes,
i.e., features having many non-zero partial correlations with other features.
These hub nodes then represent our estimate of the core set $S$. \citet{tanetal}
propose a method called {\em hglasso} for learning graphical models with hubs,
which produces a proper (non-negative definite) estimate of the inverse
covariance matrix. Their procedure uses an ADMM algorithm having
computational complexity $O(p^3)$ per iteration, which
in our experience is too slow for problems with $p=1000$ or greater.
We instead use the ``edge-out'' method of
\citet{friedman10:_applic_of_lasso_and_group}, which has complexity
$O(\min(np^2+snp,sp^2))$ per iteration. A comparison of these methods is presented in Section \ref{sec:hubmethodcomparison}.

\subsection{The edge-out procedure}
\label{sec:edgeout}
To estimate $\bB$ in step 1 of the hubNet procedure, we use the
edge-out estimator
\begin{eqnarray}
\hat{\bB}_{eo}
=\argmin_{\bB \in \real^{p \times p}:\,\bB_{ii}=0\,\forall i}
\frac{1}{2}\|\bX-\bX \bB\|^2_F+\theta\cdot\left(\gamma\|\bB_{i,.}||_1+
(1-\gamma)\sqrt{p-1}\sum_{i=1}^p \|\bB_{i,.}\|_2\right).
\label{eqn:edgeout}
\end{eqnarray}
Here, $\theta,\gamma>0$ are tuning parameters, $\|\cdot\|_F$ denotes the
Frobenius norm, and $\bB_{i,.}$ denotes the $i$th row of $\bB$.

By constraining the diagonal entries of $\bB$ to 0, the edge-out estimator
simultaneously regresses each feature onto the remaining features of $\bX$. The
procedure applies a combined $\ell_1/\ell_2$ penalty on the regression
coefficients, where the $\ell_2$ penalty encourages
zeroing-out of entire rows of $\bB$ and the $\ell_1$ penalty encourages
additional sparsity in the non-zero rows. (The original hubNet proposal of
\citet{friedman10:_applic_of_lasso_and_group} used only the $\ell_2$ penalty.)
The estimate $\hat{\bB}_{eo}$ is not symmetric. We expect the ``hub'' features
in the core set $S$ to correspond to the rows of $\bB$ having many non-zero
entries, and hence the row sums should give higher weight to these features in
steps 2 and 3. Our procedure for minimization of the edge-out objective is outlined in
Appendix \ref{app:optimization}.

 \begin{figure}
 \begin{center}
 \includegraphics[width=\textwidth]{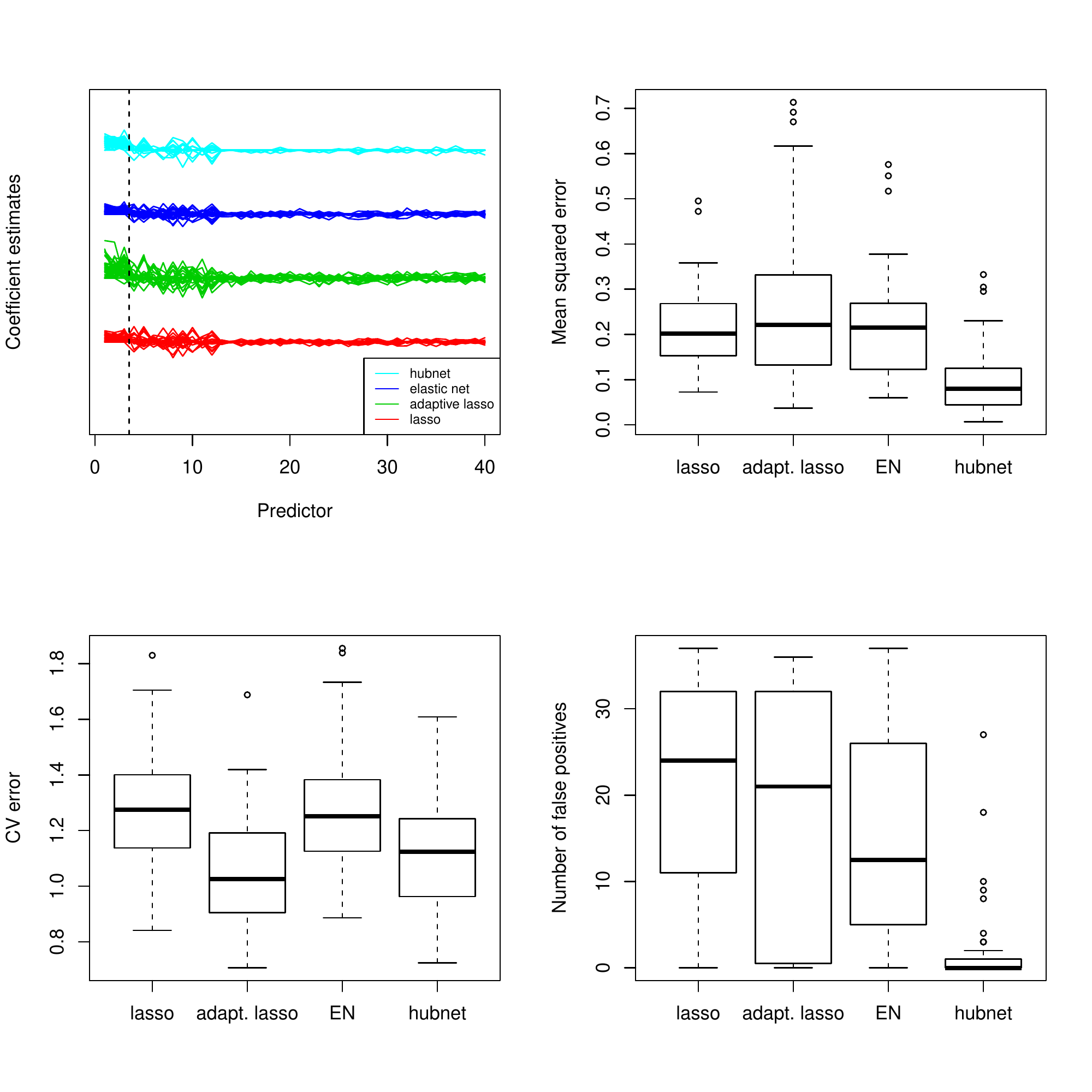}
 \end{center}

\caption[fig:fig1]{\em Estimates from 20 simulations from underlying hub model;
$n=60, p=40$, and first 3 predictors are hub  predictors and contain the signal.
The top left panel shows the estimated coefficients over 20 realizations. The
top right panel displays the mean-squared test error with the tuning parameter
chosen by cross-validation for each method. The bottom left panel shows the
minimum CV error for each realization: note that the adaptive lasso CV error is
not a valid estimate of error since the weights are estimated in a supervised manner. The bottom right panel shows the number of false positive predictors, in the smallest model where in the procedure has ``screened'', i.e. contains all of the true predictors.}
  \label{fig:fig1}
 \end{figure}

 \begin{figure}
 \begin{center}
 \includegraphics[width=\textwidth]{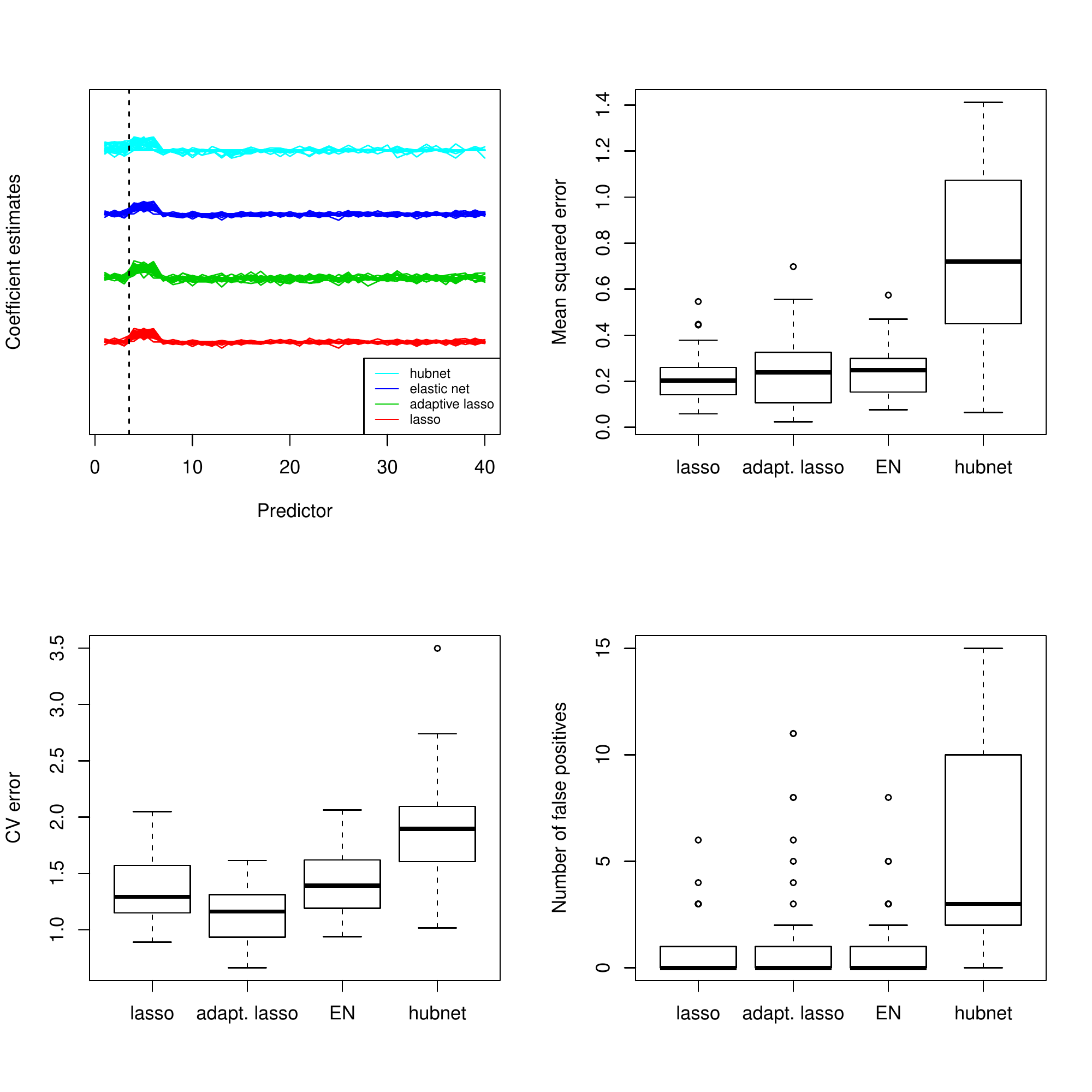}
 \end{center}
 \caption[fig:fig2]{\em Estimates from 20 simulations from underlying hub model; $n=60, p=40$, first 3 predictors are hub  predictors, but signal is a function of predictors 4 to 6. See previous figure caption for details of panels.}
 \label{fig:fig2}
 \end{figure}

\subsection{Choosing tuning parameters for edge-out}
We have two proposals for setting
the tuning parameter $\theta$ in the edge-out method.
The first is $K$-fold cross-validation, applied to the objective function  $\frac{1}{2} ||\bX-\bX \bB ||^2_F$. 
The second uses a form of generalized cross validation
\[\operatorname{GCV}(\hat{\bX})
=\frac{||\bX - \hat{\bX}||^2_2}{np - {\rm df}(\hat{\bX})}.\]
If there is only an $\ell_1$ penalty, we use for ${\rm df}(\hat{\bX})$ the
number of non-zero entries $|\hat{\bB}|_0$. If there is also an $\ell_2$
penalty, we propose the following adjustment based on our updating formula:
\[
{\rm df}(\hat{\bX}) =\sum_{i=1}^p  \frac{\|\hat{\bB}_{i,.}||_2}
{\|\hat{\bB}_{i,.}\|_2+\theta(1-\gamma)\sqrt{p-1}}
\|\hat{\bB}_{i,.}\|_0.
\]
Note that this is not an exact formula for degrees of freedom, but rather
a rough estimate.

\subsection{Simulated data example.}
Figure \ref{fig:fig1}
shows hubNet applied to a simulated data example. Here $n=60$, $p=40$, and the
first 3 predictors are the core set, explaining both $Y$ and the remaining 37
predictors. The estimated coefficients and various error rates of hubNet
over 20 realizations are shown, in comparison to the elastic net, adaptive
lasso, and lasso. We see that hubNet does a much better job at
recovering the true coefficients, which in turn leads to
substantially lower prediction error. In Figure \ref{fig:fig2} we have generated data from an adversarial setting  where the first 3 predictors are hub predictors,
but the signal is a function of predictors 4 to 6. As expected, the hubNet
procedure does poorly; however, its CV error is also high, so this poor behavior
would be detectable in practice.

\subsection{Extension to generalized regression models}
\label{sec:glm}
The hubNet procedure can be extended in  a straightforward manner to the class of generalized linear models
and other settings such as Cox's proportional hazards model.
If the outcome $Y$ depends on a parameter vector $\eta$, we assume that a core set of predictors $S$ determines both $\eta$ and the
other predictors:
\begin{eqnarray}
\eta &= &\beta_0+\bX_S\beta + \epsilon\cr
X_j &=& \bX_S\Gamma_j + \epsilon_j,\;\;j \notin S
\label{eqn:model3}
\end{eqnarray}
As in the linear case, we fit a model $\bX=\bX\bB$ using the edge-out procedure, and use
the absolute row sums of $\hat\bB$ as predictor weights in an
$\ell_1$-regularized (generalized) regression of $Y$ on $X$.

For logistic regression, an alternative strategy would assume that a model of
the form $X_j=\bX_S\Gamma^k_j + \epsilon^k_j$ for $j \notin S$ holds within
each class $k=1,2$. We may then estimate a hub model from the
{\em pooled within class} covariance matrix of $X$, and use the absolute row sums as predictor weights.
\section{Simulation studies}
\label{sec:sim}
\subsection{Comparison between hubNet, lasso and other methods}
We compare performance under different settings  between four methods: hubNet,
lasso, elastic net, and the adaptive lasso with weights set to the
inverse absolute values of the univariate regression coefficients. We
experimented with  the following four scenarios:
\begin{description}
\item{(a)} A favorable model:
\begin{align*}
Y&=\bX_S\beta + \epsilon,\;\beta = \bold{1},\;\epsilon \sim N(0,1)\\
X_j&=\bX_S\Gamma_j+\epsilon_j,\;j \in T,\;\Gamma_{ij} \sim N(0,4),
\;\epsilon_j \sim N(0,1)\\
X_j&=\epsilon_j,\;j \notin T,\;\epsilon_j \sim N(0,1)
\end{align*}
The set $S$ contains the first $s$ features, and $T$ contains
20\% of the remaining features. Hence the model (\ref{eqn:model}) is
correct but with only 20\% of non-core features depending on $\bX_S$.

\item{(b)} An adversarial model:
\begin{align*}
Y &=\bX_{S_1}\beta + \epsilon,\;\beta = \bold{1},\;\epsilon \sim N(0,1)\\
X_j &=\bX_{S_2}\Gamma_j + \epsilon_j,\; j \in T,\;\Gamma_{ij} \sim
N(0,0.25),\;\epsilon_j \sim N(0,1)\\
X_j &=\epsilon_j,\;j \notin S_2 \cup T
\end{align*}
$S_2$ contains the first $s$ features and $T$ contains 20\% of the remaining features, of which $s$ belong to $S_1$.
Hence a core set $S_2$ influences $T$, but $Y$ is explained directly by certain features in $T$ rather than $\bX_{S_2}$.
 
\item{(c)} An extreme adversarial model:
\begin{align*}
Y &=\bX_{S_1}\beta + \epsilon,\;\beta = \bold{1},\;\epsilon \sim N(0,1)\\
X_j &=\bX_{S_2}\Gamma_j + \epsilon_j,\;j \notin S_2,
\;\Gamma_{ij} \sim N(0,0.25),\;\epsilon_j \sim N(0,1)\\
X_j &=\epsilon_j,\;j \in S_2
\end{align*}
$S_2$ contains the first $s$ features and $S_1$ contains the next $s$ features.
This setup is the same as in (b) above, except $T$ is now the set of all features outside $S_2$.
 
\item{(d)} A neutral model:
\begin{align*}
Y &=\bX_S\beta+\epsilon,\;\beta=\bold{1},\;\epsilon \sim N(0,1)\\
X &\sim N(0, \bSigma)
\end{align*} 
$S$ contains the first $s$ features, and $\bSigma$ is a random
positive-definite covariance matrix (generated 
using the R function {\tt genPositiveDefMat}) with the ratio of largest
to smallest eigenvalue set to 10.
\end{description}
For each scenario, we consider $(n, p, s) = (100, 500, 10)$ and
$(200, 1000, 20)$, and we also scale each feature to have variance 1 before
applying each of the four methods. For hubNet, the edge-out tuning parameter
$\theta$ is set by minimizing GCV, and we fix $\gamma=1/2$. For the elastic net,
we also fix $\alpha=1/2$. The main tuning parameter $\lambda$ in all four
methods (corresponding to the tuning parameter for the adaptive lasso step in hubNet) is set by 10-fold cross-validation.
 
We evaluate performance using the proportion of falsely detected features 
(FP), the proportion of true features that are undetected (FN),
the cross-validation mean square prediction error in the training set (cvm), mean square prediction error in
the test set, and the total number of selected features.
A summary of these values averaged across 100 repetitions of each
scenario is presented in Tables \ref{tab:tab1} to \ref{tab:tab4}, with standard deviations reported for cvm and test error.
 \begin{table}[H]
\centering
\caption{Comparison of hubNet with other methods in scenario (a)}
\label{tab:tab1}
\begin{adjustbox}{width=0.6\textwidth}
\begin{tabular}{lllllll}
  \hline
    $(n,p,s) = (100, 500, 10)$ &&&&  &  \\ 
  \hline
 & cvm(se) & FN & FP & features & test.error(se) \\ 
  \hline
llasso & 1.557(0.234) & 0.940 & 0.973 & 30.120 & 1.623(0.322) \\ 
  elasticNet & 1.568(0.249) & 0.904 & 0.973 & 39.230 & 1.630(0.348) \\ 
  adaptiveLasso & 1.486(0.257) & 0.966 & 0.970 & 11.300 & 1.583(0.332) \\ 
  hubNet & 1.208(0.173) & 0.004 & 0.278 & 16.580 & 1.335(0.215) \\ 
  \hline
    $(n,p,s) = (200, 1000, 20)$ &&&&  &  \\ 
     \hline
     & cvm(se) & FN & FP & features & test.error \\ 
  \hline
lasso & 1.556(0.210) & 0.934 & 0.977 & 59.540 & 1.564(0.211) \\ 
  elasticNet & 1.576(0.219) & 0.901 & 0.971 & 71.360 & 1.571(0.215) \\ 
  adaptiveLasso & 1.554(0.258) & 0.960 & 0.963 & 20.860 & 1.613(0.311) \\ 
  hubNet & 1.184(0.131) & 0.003 & 0.262 & 29.330 & 1.278(0.143) \\ 
\hline
\end{tabular}
\end{adjustbox}
\end{table}

 \begin{table}[H]
\centering
\caption{Comparison of   hubNet with other methods in scenario (b)}
\label{tab:tab2}
\begin{adjustbox}{width=0.6\textwidth}
\begin{tabular}{llllll}
  \hline
    $(n,p,s) = (100, 500, 10)$ &&&&&   \\ 
  \hline
& cvm(se) & FN & FP & features & test.error(se)  \\ 
   \hline
lasso & 5.479(2.233) & 0.032 & 0.847 & 66.330 & 4.588(2.239)\\ 
  elasticNet & 7.017(2.156)& 0.052 & 0.863 & 72.940 & 6.140(2.563) \\ 
  adaptiveLasso & 4.878(1.773)& 0.162 & 0.786 & 41.650 & 5.867(2.623) \\ 
  hubNet & 3.891(1.524) & 0.012& 0.784 & 47.880 & 3.373(1.484) \\ 
\hline
    $(n,p,s) = (200, 1000, 20)$ &&&&&   \\ 
     \hline
 & cvm(se) & FN & FP & features & test.error(se)  \\ 
   \hline
lasso & 15.277(4.159)& 0.128 & 0.854 & 126.800 & 12.611(5.519) \\ 
  elasticNet & 17.328(3.555) & 0.150 & 0.858 & 126.910 & 15.485(4.567) \\ 
  adaptiveLasso & 12.125(2.537)& 0.224 & 0.758 & 67.570 & 13.183(3.658)  \\ 
  hubNet & 7.218(3.686) & 0.020 & 0.717 & 72.450 & 6.181(3.262) \\ 
\hline
\end{tabular}
\end{adjustbox}
\end{table}
 \begin{table}[H]
\centering
\caption{Comparison of   hubNet with other methods in scenario (c)}
\label{tab:tab3}
\begin{adjustbox}{width=0.6\textwidth}
\begin{tabular}{llllll}
  \hline
    $(n,p,s) = (100, 500, 10)$ &&&&&    \\ 
  \hline
 & cvm(se) & FN & FP & features & test.error(se)  \\ 
  \hline
lasso & 2.619 (0.820) & 0.001 & 0.817 & 57.680 & 2.531(0.807) \\ 
  elasticNet & 3.530(1.183) & 0.000 & 0.856 & 71.890 & 3.143(0.984) \\ 
  adaptiveLasso & 5.988(1.889) & 0.193 & 0.786 & 40.860 & 6.258(2.086) \\ 
  hubNet & 5.875(2.296) & 0.137 & 0.546 & 19.170 & 5.788(2.693) \\ 
\hline
    $(n,p,s) = (200, 1000, 20)$ &&&&&    \\ 
     \hline
   & cvm(se) & FN & FP & features & test.error(se)  \\ 
  \hline
lasso & 2.776(0.525)&0.000 & 0.767 & 86.720 & 2.866(0.642) \\ 
  elasticNet & 3.915(0.809) & 0.000 & 0.798 & 99.710 & 3.664(0.877) \\ 
  adaptiveLasso & 13.466 (2.344) & 0.243 & 0.796 & 77.100 & 13.135(2.883) \\ 
  hubNet & 22.007(4.359) & 0.823 & 0.878 & 22.490 & 21.875(4.600) \\ 
\hline
\end{tabular}
\end{adjustbox}
\end{table}
 \begin{table}[H]
\centering
\caption{Comparison of hubNet with other methods in scenario (d)}
\label{tab:tab4}
\begin{adjustbox}{width=0.6\textwidth}
\begin{tabular}{llllll}
  \hline
    $(n,p,s) = (100, 500, 10)$ &&&& &  \\ 
  \hline
 & cvm(se) & FN & FP & features & test.error(se) \\ 
  \hline
lasso & 2.486(0.514) & 0.000 & 0.800 & 54.210 & 2.683(0.778)\\ 
  elasticNet & 3.948(1.110) & 0.000 & 0.850 & 69.600 & 3.649(1.322) \\ 
  adaptiveLasso & 2.038(1.631) & 0.012 & 0.703 & 37.960 & 3.085(2.723) \\ 
  hubNet & 1.709(0.354) & 0.000 & 0.719 & 38.710 & 2.156(0.617) \\ 
\hline
    $(n,p,s) = (200, 1000, 20)$ &&&& &  \\ 
     \hline
     & cvm(se) & FN & FP & features & test.error(se) \\ 
  \hline
lasso & 2.380(0.364) & 0.000 & 0.801 & 104.400 & 2.668(0.623) \\ 
  elasticNet & 3.374(0.694) & 0.000 & 0.839 & 126.780 & 3.317(0.888) \\ 
  adaptiveLasso & 3.475(1.824) & 0.017 & 0.488 & 41.740 & 4.615(2.687) \\ 
  hubNet & 1.641(0.205)  & 0.000 & 0.689 & 66.120 & 2.131(0.415) \\ 
\hline
\end{tabular}
\end{adjustbox}
\end{table}

HubNet outperforms the other three methods in scenario (a) as expected.
Perhaps surprisingly, it also seems to outperform the other methods under scenarios (b) and (d). In the extreme adversarial scenario (c), hubNet performs worse than
the other methods, although this can be detected in cross-validation.

In Figure \ref{fig:paths} of Appendix \ref{app:comparison},
we track FP and FN along the solution paths of the various methods as $\lambda$
varies. The results are in line with the above.

\section{Application to real datasets}
\label{sec:real}
We compare hubNet with the lasso and elastic net on three real data examples.
The following table summarizes the cross-validation errors, test errors,
number of selected features, and number of such features in common with those
selected by lasso.
 \begin{table}[H]
\centering
\caption{\em Comparisons among lasso, elasticNet and hubNet on three
real data sets.}
\label{tab:realdata}
\begin{adjustbox}{width=0.8\textwidth}
\begin{tabular}{l|l|llll}
\hline
&& cvm(se) &Num. features & test error  & common features (lasso) \\\hline
 Breast Cancer Data&lasso& $5.15\%(3.86\%)$ & 46 & -- & -- \\
 $p =806 $ &elasticNet& $5.85\%(3.97\%)$ & 303 & --& 46 \\
  $n_{\text{train}}= 15359$&hubNet& $3.52\%(2.92\%)$ &92& -- & 26\\\hline
   && cvm(se) & Num. features & test p-value  &common features (lasso) \\\hline
  Kidney Cancer Data&lasso& 9.89(0.56) & 20 & 0.294 & -- \\
  $p =14814 $&elasticNet& 9.96(0.56)  &11 & 0.125 & 9\\
  $n_{\text{train}}= 88, n_{\text{test}}=89$&hubNet&9.99(0.42) & 1 & 0.008& 0\\\hline
   &&cvm(se)& Num. features & test p-value  &common features (lasso)\\\hline
  DLBCL-patient Data&lasso& 10.9(0.39) & 29&0.076 & -- \\
   $p =7399 $&elasticNet& 10.9(0.39)& 37 & 0.052 & 28 \\
  $n_{\text{train}}= 156, n_{\text{test}}=79$&hubNet& 11.0(0.24)& 2& 0.035  & 0 \\\hline
\end{tabular}
\end{adjustbox}
\end{table}

\subsection*{Example: Lipidomic breast cancer data}
\begin{figure}[H]
\begin{center}
\includegraphics[width=.5\textwidth]{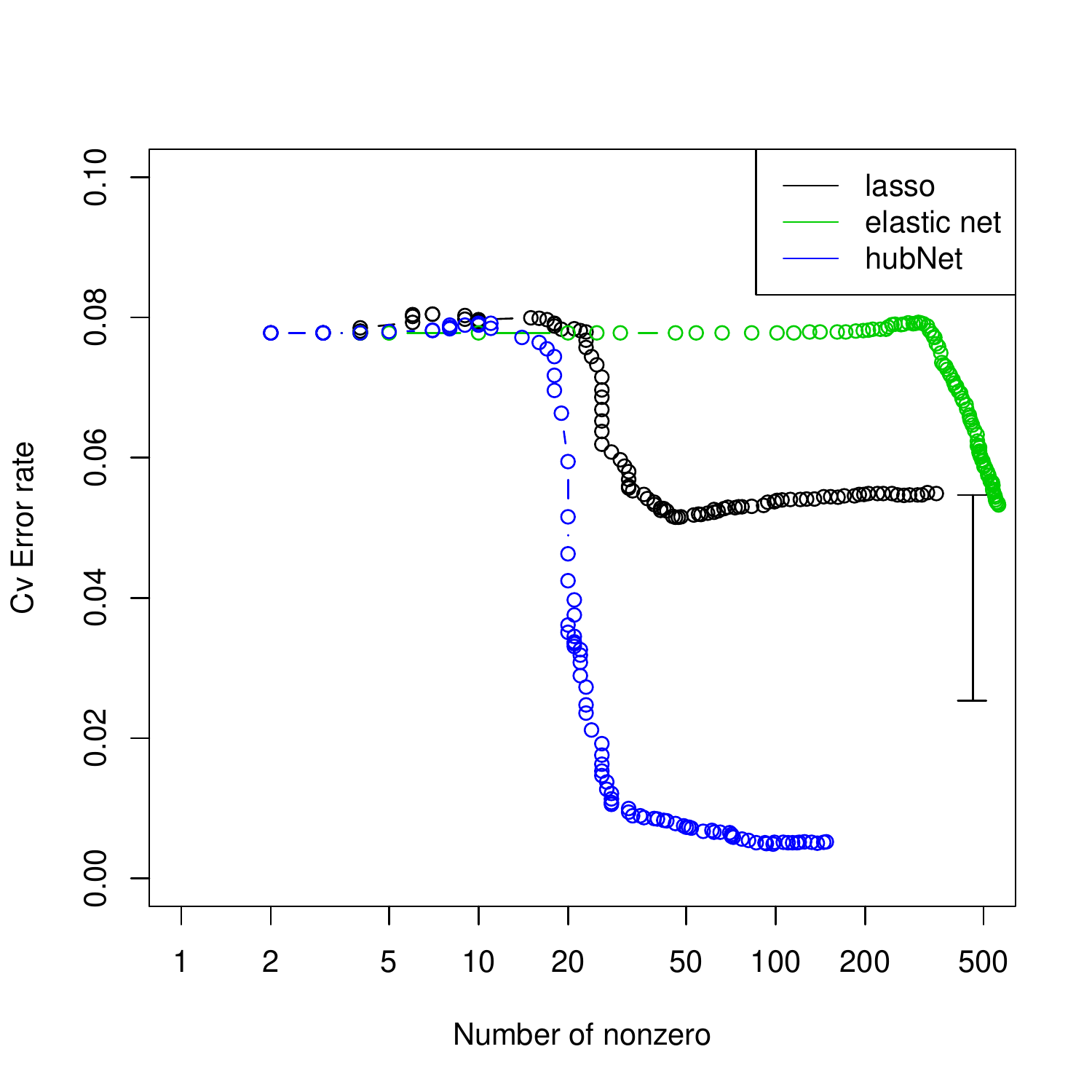}
\end{center}
\caption[fig:breast-hubNet]{\em Breast cancer data classification error rates}
\label{fig:breast-hubNet}
\end{figure}

This data, from the lab of RT's collaborator Livia Schiavinato Eberlin at UT
Austin, consists of 806 features measured on 15,359 pixels in tissue
images from 24 breast cancer patients. The pixels are divided into two classes,
normal and cancer, and we fit a regularized logistic regression model using
each procedure. Cross-validation classification errors are shown in
Figure \ref{fig:breast-hubNet} as $\lambda$ varies.
Table \ref{tab:realdata} reports results for $\lambda$ selected using
5-fold cross-validation.
\subsection*{Example: B cell lymphoma gene expression data}
This data from \citet{aR02} consists of survival times (observed or
right-censored) and 7399 gene expression features for 240 patients with
diffuse large B-cell lymphoma (DLBCL). We divided the data with survival
time $Y>0$ into 156 training and 79 test samples, and trained a regularized
proportional hazards model using each procedure. The
p-value of the log-likelihood ratio (LR) statistic of this trained model evaluated on the test set
is shown in the left subplot of Figure \ref{fig:BrookBcell} as $\lambda$
varies. Table \ref{tab:realdata} reports results
for $\lambda$ selected using 20-fold cross-validation.

\subsection*{Example: Kidney cancer gene expression data}
\begin{figure}[H]
\begin{center}
\includegraphics[width=0.9\textwidth]{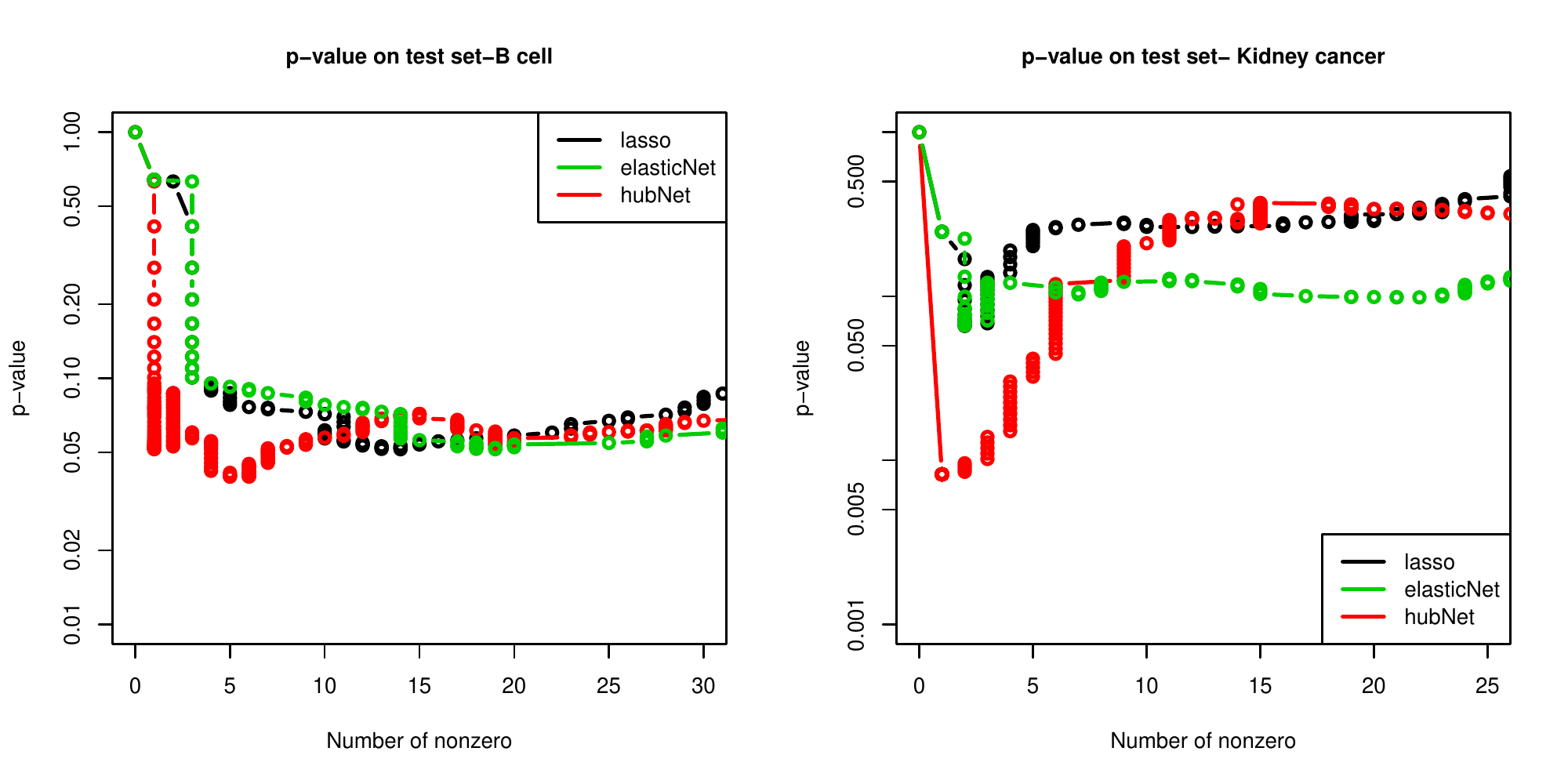}
\end{center}
\caption[fig:Bcell]{\em Results for B-cell lymphoma (left) and kidney
cancer (right): p-values of LR statistics}
\label{fig:BrookBcell}
\end{figure}

This data from \citet{Zhaoetal2005} consists of survival times and 14,814 gene
expression features for 177 patients with conventional renal cell carcinoma.
We divided the data into 88 training samples and 89 test samples and trained a
regularized proportional hazards model using each procedure.
For computational reasons, hubNet was fit using the 7999 features with largest
absolute row sum in the pairwise correlation matrix; lasso and elastic net were
fit using all features. Test set LR p-values are shown in the right
subplot of Figure \ref{fig:BrookBcell} as $\lambda$ varies, and Table
\ref{tab:realdata} reports results for $\lambda$ selected using
8-fold cross validation.

\section{Theory}
\label{sec:theory}
In this section, we study recovery of the core set $S$ assuming that our generating model
(\ref{eqn:linearmodel}, \ref{eqn:model}) holds. We first establish conditions
under which the unsupervised edge-out procedure alone can recover $S$, and then discuss recovery of $S$ by the second adaptive lasso step even if the edge-out procedure does not yield perfect recovery.

We assume the asymptotic regime $n,p \to \infty$ where $s \ll \min(n,p)$,
as well as a fully random design where the rows of $\bX$
are independent and distributed as $N(0,\bSigma)$, normalized so that
$\bSigma_{jj}=1$ for all $j=1,\ldots,p$. Without loss of
generality, we suppose $S$ contains the first $s$ predictors.
By (\ref{eqn:model}),
if $X:=(X_S,X_{S^C}) \sim N(0,\bSigma)$, then
\begin{align}
X_S &\sim N(0,\bSigma_{SS}),\cr
X_j|X_S &\overset{ind}{\sim} N(X_S^T\Gamma_j,\sigma_j^2),\;\;j \in S^C
\label{sec:basicmodel}
\end{align}
where $\sigma_j^2=\Var(\epsilon_j) \in (0,1)$. Specifically,
$\bGamma:=(\Gamma_{s+1},\ldots,\Gamma_p)$ is given by
$\bSigma_{SS}^{-1}\bSigma_{SS^C}$.
We assume that this model holds in all of the results that follow.



\subsection{Recovery of the core set using the edge-out procedure}
We analyze recovery of $S$ by the edge-out procedure
applied with only the group-lasso penalty term in (\ref{eqn:edgeout}),
corresponding to the setting $\gamma=0$. For any matrix $\bM$, denote by
$\bM_{i,.}$ and $\bM_{.,j}$ the $i$th row and $j$th column of $\bM$. We use
the following operator norms which measure the maximum $\ell_1$ and $\ell_2$
norm of any row of $\bM$:
\[\|\bM\|_\infty:=\sup_{\|x\|_\infty=1}\|\bM x\|_\infty=\max_i
\|\bM_{i,.}\|_1,\hspace{0.5in}
\|\bM\|_{\infty,2}:=\sup_{\|x\|_2=1} \|\bM x\|_\infty=\max_i
\|\bM_{i,\cdot}\|_2.\]
We define also the usual spectral norm, given by the largest singular value of
$\bM$,
\[\|\bM\|_2:=\sup_{\|x\|_2=1} \|\bM x\|_2=\sigma_{\max}(\bM).\]

We show that in the asymptotic regime $n,p \to \infty$, the edge-out procedure
can recover
the true core set $S$ for a suitable choice of the tuning parameter $\theta$
when the following conditions hold:
\begin{assumptions}\label{assump:Cmin}
Let $\lambda_{\min}(\bSigma_{SS})$ be the smallest eigenvalue of
$\bSigma_{SS}$. For a fixed constant $C_{\min}>0$,
$\lambda_{\min}(\bSigma_{SS}) \geq C_{\min}$.
\end{assumptions}
\begin{assumptions}\label{assump:irrepresentable}
Define $\bD:=\operatorname{diag}(1/\|\Gamma_{s+1}\|_2,\ldots,1/\|\Gamma_p\|_2)$.
For a fixed constant $\delta \in (0,1]$,
\[\|\bGamma^T\bD\bGamma\|_{\infty,2} \leq 1-\delta.\]
\end{assumptions}
\begin{assumptions}\label{assump:s}
(Number of hub nodes). The size $s$ of the core set satisfies
\[s \ll \min(\sqrt{n},n/\log p).\]
\end{assumptions}
\begin{assumptions}\label{assump:Gmin}
(Hub strength). The minimum hub strength
$\Gamma_{\min}=\min_i \|\bGamma_{i,.}\|_2$ satisfies
\[\Gamma_{\min} \gg \max(\|\bGamma^T\|_\infty,1)
\|\bSigma_{SS}^{-1}\|_\infty\max(1,\sqrt{p/n},\sqrt{p\log p}/n).\]
\end{assumptions}

Under these assumptions, we can ensure perfect recovery of
the core set $S$ by the edge-out method:
\begin{restatable}{theorem}{thmA}
\label{thm:thmA}
Let $\hat{\bB}:=\hat{\bB}_{eo}$ be the edge-out estimate in (\ref{eqn:edgeout})
applied with $\gamma=0$, and denote $\hat{S}=\{i:\|\hat{\bB}_{i,.}\|_2>0\}$.
Suppose Assumptions \ref{assump:Cmin}, \ref{assump:irrepresentable},
\ref{assump:s}, and \ref{assump:Gmin} hold. Defining
$\theta_n=\theta\sqrt{p-1}/n$, if the tuning parameter $\theta$ is
chosen so that
\begin{equation}\label{eq:lambdachoice}
\frac{\Gamma_{\min}}{\max(\|\bGamma^T\|_\infty,1) \|\bSigma_{SS}^{-1}\|_\infty}
\gg \theta_n \gg \max\left(1,\sqrt{\frac{p}{n}},\frac{\sqrt{p\log p}}{n}
\right),
\end{equation}
then
\[P[\hat{S}=S] \to 1.\]
\end{restatable}

Assumption \ref{assump:Cmin} ensures that the hub features are not too
correlated. Assumptions \ref{assump:s} and \ref{assump:Gmin} restrict the
maximal
size of the core set and minimal ``strength'' of the hub features, as measured
by the minimum $\ell_2$ row norm of $\bGamma$. Let us remark that our
normalization implies an additional implicit constraint on $s$, namely
$p \geq \sum_{j \in S^C} \Var(X_j)=
\sum_{j \in S^C} \Gamma_j^T\bSigma_{SS}\Gamma_j+\sigma_j^2
\geq \|\bGamma\|_F^2C_{\min} \geq sC_{\min}\Gamma_{\min}^2$,
so by Assumption \ref{assump:Gmin}
\[s \ll \frac{\min(n,p,n^2/\log p)}{\max(\|\bGamma^T\|_\infty,1)^2
\|\bSigma_{SS}^{-1}\|_\infty^2}.\]
In the worst case, we have the upper bounds
$\|\bSigma_{SS}^{-1}\|_\infty \leq \sqrt{s}\|\bSigma_{SS}^{-1}\|_2 \leq
\sqrt{s}/C_{\min}$ and
$\|\bGamma^T\|_\infty \leq \sqrt{s}\|\bGamma^T\|_{\infty,2} \leq
\sqrt{s/C_{\min}}$,
where the latter bound follows from our normalization condition
\begin{equation}\label{eq:Gammacolbound}
\|\bGamma^T\|_{\infty,2}^2C_{\min}
\leq \max_{j \in S^C} \Gamma_j^T\bSigma_{SS}\Gamma_j \leq \Var(X_j) \leq 1.
\end{equation}
Assuming $\log p \ll \sqrt{n}$,
recovery can occur in this worst case when $s \ll \min(n^{1/3},p^{1/3})$.
In the best case where an ``irrepresentable condition''
$\|\bGamma^T\|_\infty \leq 1$ holds (see below) and $\bSigma_{SS}=\Id$,
then we have $\max(\|\bGamma^T\|_\infty,1)\|\bSigma_{SS}^{-1}\|_\infty=1$,
and recovery can occur for $s \ll \min(\sqrt{n},p)$.

Assumption \ref{assump:irrepresentable} is analogous to but much weaker
than the ``irrepresentable condition'' of \citet{zhao2006model} (see also
\citet{wainwright2009sharp}) that is required for perfect support recovery by
the standard lasso procedure. In our random design setting, the
irrepresentable condition corresponds to
\begin{equation}\label{eq:regularirrepresentable}
\|\bGamma^T\|_\infty \leq 1-\delta
\end{equation}
for some $\delta \in (0,1]$. When (\ref{eq:regularirrepresentable}) holds,
Assumption \ref{assump:irrepresentable} is
implied by $\|\bGamma^T\bD\bGamma\|_{\infty,2}
\leq \|\bGamma^T\|_\infty\|\bD\bGamma\|_{\infty,2}=\|\bGamma^T\|_\infty$.
The following example illustrates that Assumption \ref{assump:irrepresentable}
is weaker than (\ref{eq:regularirrepresentable}):
\begin{example}
Suppose the entries of $\bGamma$ are i.i.d.\ and equal to
$(1-2\delta)/\sqrt{s}$ or $-(1-2\delta)/\sqrt{s}$ each with probability 1/2.
Then $\|\bGamma^T\bD\bGamma\|_{\infty,2} \leq \|\bGamma^T\|_{\infty,2}
\|\bD\|_2\|\bGamma\|_2
=\sqrt{s/(p-s)}\|\bGamma\|_2$. If $p \to \infty$ with $s \ll
p$, the maximal singular value of $\bGamma$ satisfies,
for any fixed $\varepsilon>0$ with probability approaching 1,
$\|\bGamma\|_2 \leq (1+\varepsilon)\sqrt{p} \cdot (1-2\delta)/\sqrt{s}$.
(See e.g.\ Theorem 5.39 of \citet{vershynin}.)
Hence for large $p$, $\bGamma$ satisfies Assumption
\ref{assump:irrepresentable} with high probability.
However, $\|\bGamma^T\|_\infty=(1-2\delta)\sqrt{s} \gg 1$.
\end{example}
This example shows that Assumption \ref{assump:irrepresentable} can hold even
in the worst-case setting where $\|\bGamma^T\|_\infty \asymp \sqrt{s}$, as
long as the non-hub features are not influenced by the hub features
``in the same way''.

\subsection{Recovery of the core set using adaptive lasso}
We now consider the linear model (\ref{eqn:linearmodel}) where
$\epsilon=(\epsilon_1,\ldots,\epsilon_p)$ is independent of $\bX$ with
$\epsilon_i \overset{iid}{\sim} N(0,\sigma^2)$. We study recovery of $S$ by the
adaptive lasso step of the hubNet procedure in two cases: (a) the edge-out estimate yields exact recovery of $S$ , and (b)
it yields a superset of $S$.

Let $w_1,\ldots,w_p \in (0,\infty]$ be any feature weights
derived from $\bX$. (Setting
$w_i=\infty$ corresponds to $\|(\hat{\bB}_{eo})_{i,.}\|_2=0$,
i.e.\ a hard constraint that requires $\beta_i=0$.) Define
\[\rho:=w_{\max}(S)/w_{\min}(S^C),\;\;\;\;
w_{\min}(S^c):=\min_{i\in S^c} w_i,\;\;\;\;w_{\max}(S):=\max_{i\in S} w_i,\]
with the convention $\infty/\infty=\infty$. We consider the following conditions as $n,p \to \infty$:
\begin{assumptions}\label{assump:wgap}
There exists $\eta \in (0,1]$ such that with probability approaching 1,
\[\rho\sqrt{\frac{s}{C_{\min}}}\left(1+\sqrt{\frac{12\log p}{n}}\right)
\leq 1-\eta.\]
\end{assumptions}
\begin{assumptions}\label{assump:betamin}
The minimum predictor strength
$\beta_{\min}=\min_{i \in S} |\beta^*_i|$ satisfies
\[\beta_{\min} \gg \sigma\sqrt{\frac{s\log p}{n}
\left(1+\frac{\log p}{n}\right)}.\]
\end{assumptions}
Then, under our model (\ref{eqn:linearmodel}) and (\ref{eqn:model}),
the following result holds for the adaptive lasso:
\begin{theorem}\label{thm:thmB}
Let $n,p \to \infty$ such that $s \ll n$ and Assumption \ref{assump:Cmin}
holds. Furthermore, let
$w_1,\ldots,w_p \in (0,\infty]$ be weights (depending on $\bX$)
such that Assumption \ref{assump:wgap} holds. Denote by
$\hat{\beta}_0,\hat{\beta}$ the estimator minimizing the adaptive lasso
objective (\ref{eq:adaptivelasso}), and let $\hat{S}=\{i:\hat{\beta}_i \neq
0\}$.
\begin{description}
\item {(a)} Denoting $\lambda_n=\lambda/n$, if the tuning parameter $\lambda$ of the adaptive lasso is 
chosen such that
\[\lambda_n \gg \frac{1}{w_{\min}(S^C)}
\sigma\sqrt{\frac{\log p}{n}\left(1+\frac{\log p}{n}\right)}\]
with probability approaching 1, then
\[P[\hat{S} \subseteq S] \to 1.\]
\item{(b)} If, in addition, Assumption \ref{assump:betamin} holds and
$\lambda_n \ll \beta_{\min}/(w_{\max}(S)\sqrt{s})$ with probability approaching
1, then
\[P(\hat{S}=S) \to 1.\]
\end{description}
\end{theorem}

This result holds for any procedure that selects $w_1,\ldots,w_p$ using $\bX$. Assumption \ref{assump:betamin} is comparable to the beta-min condition in Theorem 3 of \citet{wainwright2009sharp} for the standard lasso procedure, if $\sqrt{s}$ is replaced by $\|\bSigma_{SS}^{-1/2}\|_\infty^2$. In the context of hubNet, Assumption \ref{assump:wgap} should be interpreted as a weakening of the conditions required for selection consistency of $S$ by the edge-out procedure alone: If the edge-out procedure successfully recovers
$S$, then $w_{\min}(S^c)=\infty$ and $w_{\max}(S)<\infty$, so Assumption \ref{assump:wgap} holds. More generally, Assumption
\ref{assump:wgap} holds when there is a separation in size
between the rows of $\hat{\bB}_{eo}$ belonging to $S$ and to $S^C$,
even if the rows belonging to $S^C$ are not identically 0.

We prove Theorems \ref{thm:thmA} and \ref{thm:thmB} in Appendix
\ref{app:proof}. The proof of Theorem \ref{thm:thmB} is a simple application of the Sign Recovery Lemma in \citet{zhou2009adaptive} for the adaptive lasso procedure. A more refined statement of Theorem \ref{thm:thmB} in terms of the quantities $\|\bGamma^T\|_\infty$ and $\|\bSigma_{SS}^{-1}\|_\infty$, similar to that of Theorem \ref{thm:thmA}, is possible, although we have stated the above version for simplicity and interpretability.

\section{Further topics}
\label{sec:further}
\subsection{Adaptive, non-linear models}
We can extend our basic model (\ref{eqn:model}) to allow the dependence of $Y$ on the core set of predictors to be of a more general form:
\begin{eqnarray}
Y &= &f(\bX_S)+ \epsilon\\
X_j &=& \bX_S\Gamma_j + \epsilon_j,  j \notin S
\label{eqn:model4}
\end{eqnarray}
Here $f(\cdot)$ is a general, non-linear function.
For this model, we can estimate hub weights $s_j$ as before and then apply a more flexible prediction procedure such as  random forests or gradient boosting
using the $s_j$ as feature weights.   With random forests, the candidate predictors for splitting
are chosen at random. Hence it is natural to implement feature weighting  by  using the weights to determine  the probabilities in this sampling.
For example, the {\tt ranger} package in R provides this option.

\begin{figure}
\begin{center}
\includegraphics[width=.5\textwidth]{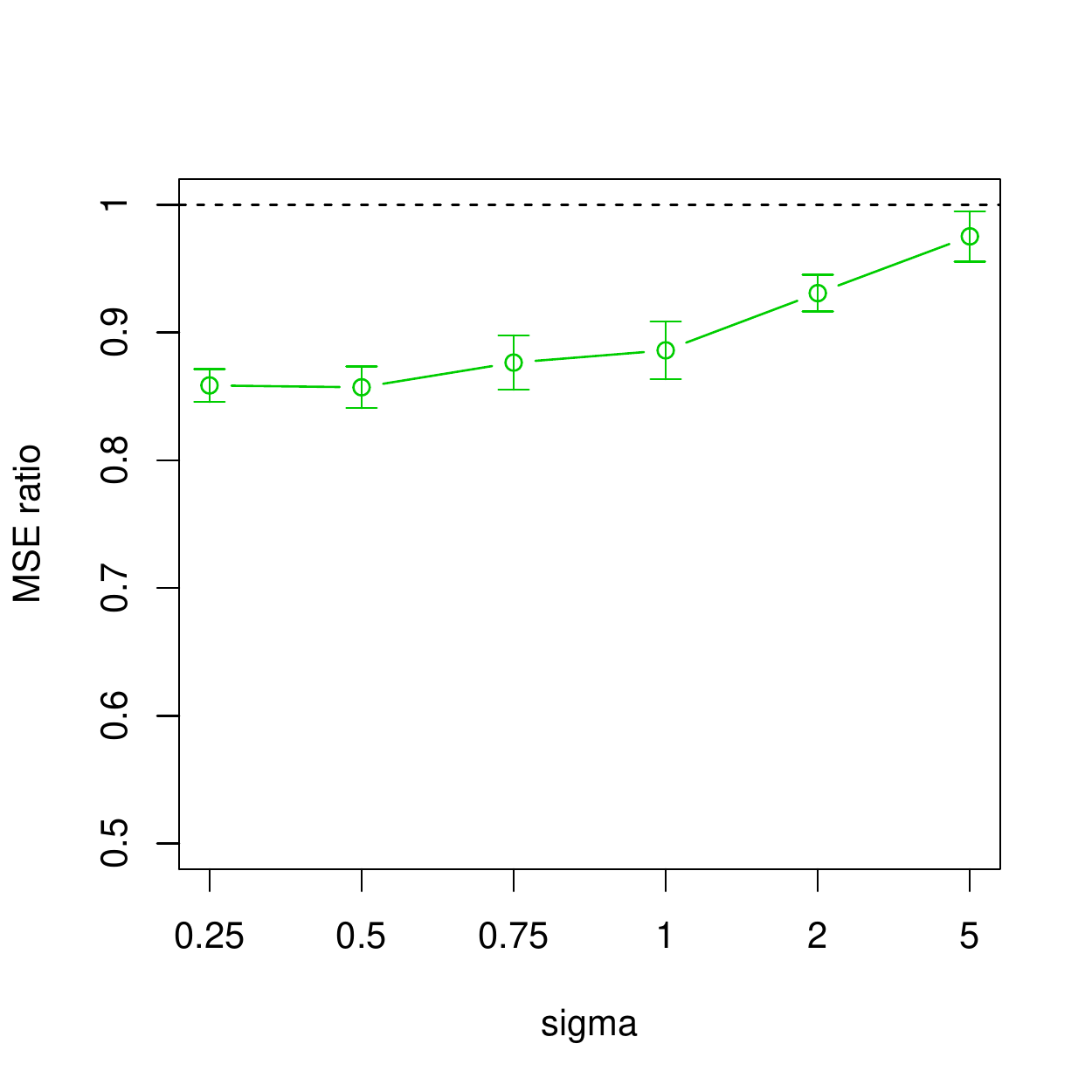}
\end{center}
\caption{\em MSE ratio of the hub-weighted random forest to the standard random
forest, for varying error standard deviation}
\label{fig:ranfor}
\end{figure}

We tried this idea in the example of Figure \ref{fig:fig1}, with additional interactions $.5x_1 x_2$ and $-2x_2 x_3$  added to the mean of $Y$, so that there were
interactions for the random forest to find. We used sampling probabilities  proportional to $s_j^2$.  In Figure \ref{fig:ranfor} we show the ratio of the
mean squared error of the hubNet/RF over that for the vanilla random forest, as the error standard deviation $\sigma$ is varied.
We see that the hub weights can decrease the mean squared error by as much as 15\%.

\subsection{Random forests: a drug discovery application}
\begin{figure}
\begin{center}
\includegraphics[width=.75\textwidth]{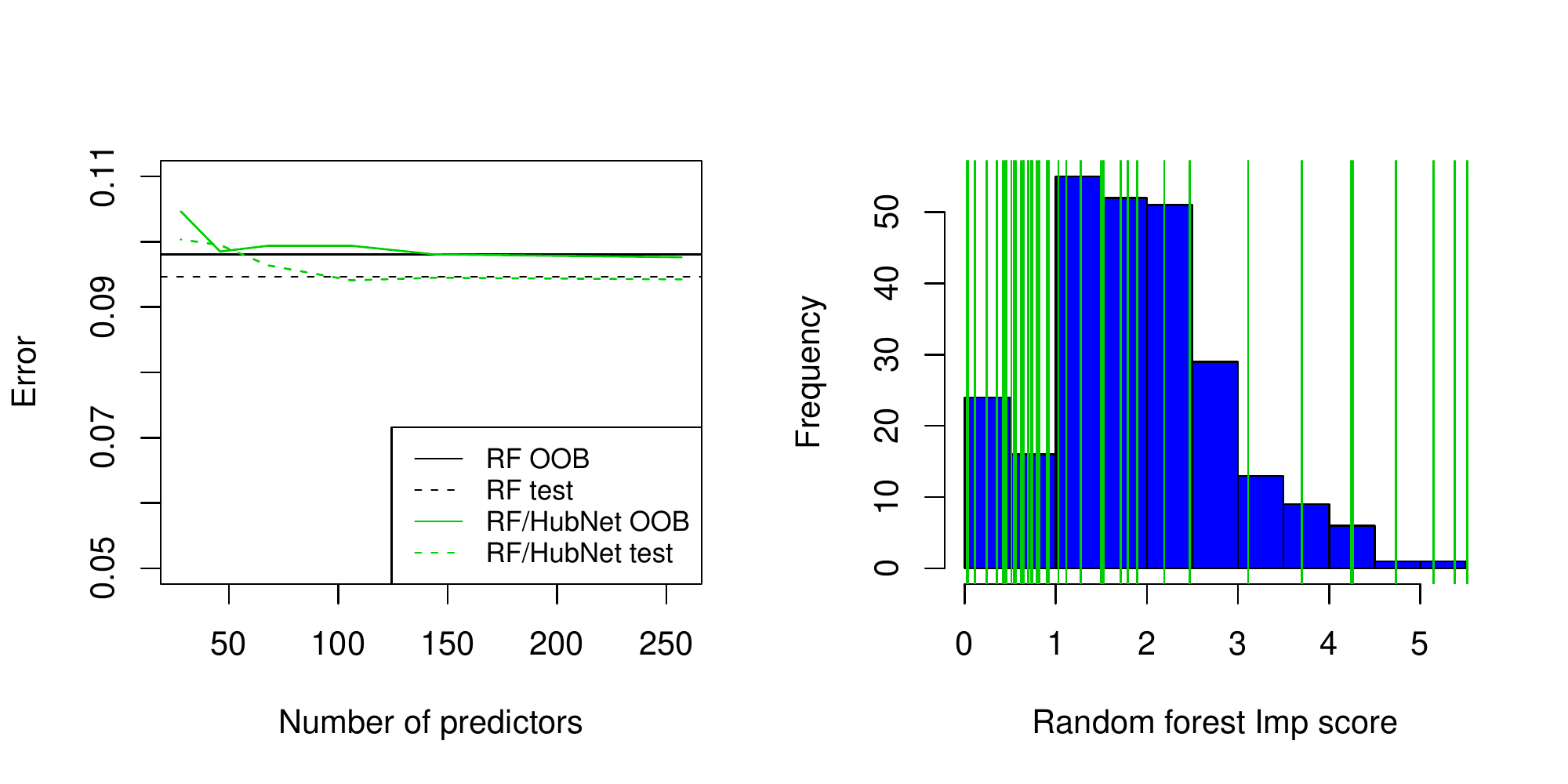}
\end{center}
\caption[fig:toxRF]{\em Results for drug discovery dataset. Left panel show out-of-bag error and test error for vanilla random forest (horizontal lines), and the same
for hubNet/RF as a function of the number of  features having non-zero hub weights (by varying $\theta$ in the edge-out model).
We see that the error increases very little, even as the number of  number of features is reduced to about one-tenth  (28) of the total  number.
These 28 features are indicated by the green lines in the right panel, superimposed on the RF impurity importance scores for all features. 
}
\label{fig:toxRF}
\end{figure}

We consider classification data collected by the NCI, described in  \citet{feng2003} and analyzed further in \citet{chipman2010}.  It consists of 
$p= 266$ molecular characteristics of $n = 29,374$
compounds, of which $542 $ were classified as active ($Y=1$). These predictors represent
topological aspects of molecular structure.  We randomly created  training and test sets of equal size, and for computational reasons we downsampled the class 0 
cases to a set of size  2000 out of  the 14,687  class 0s in the training set.
We applied both random forests and hubNet/RF, using the {\tt ranger} package in R.
The results in Figure \ref{fig:toxRF} show  that the hubNet weighting can reduce the number of features by a factor of about 10 (down to 28) with barely any loss in accuracy, and these 28 features
would not be detectable from standard RF importance scores (right panel).

\subsection{Post-selection inference}
\label{sec:postsel}
\begin{figure}[ht]
\begin{center}
\includegraphics[width=.75\textwidth]{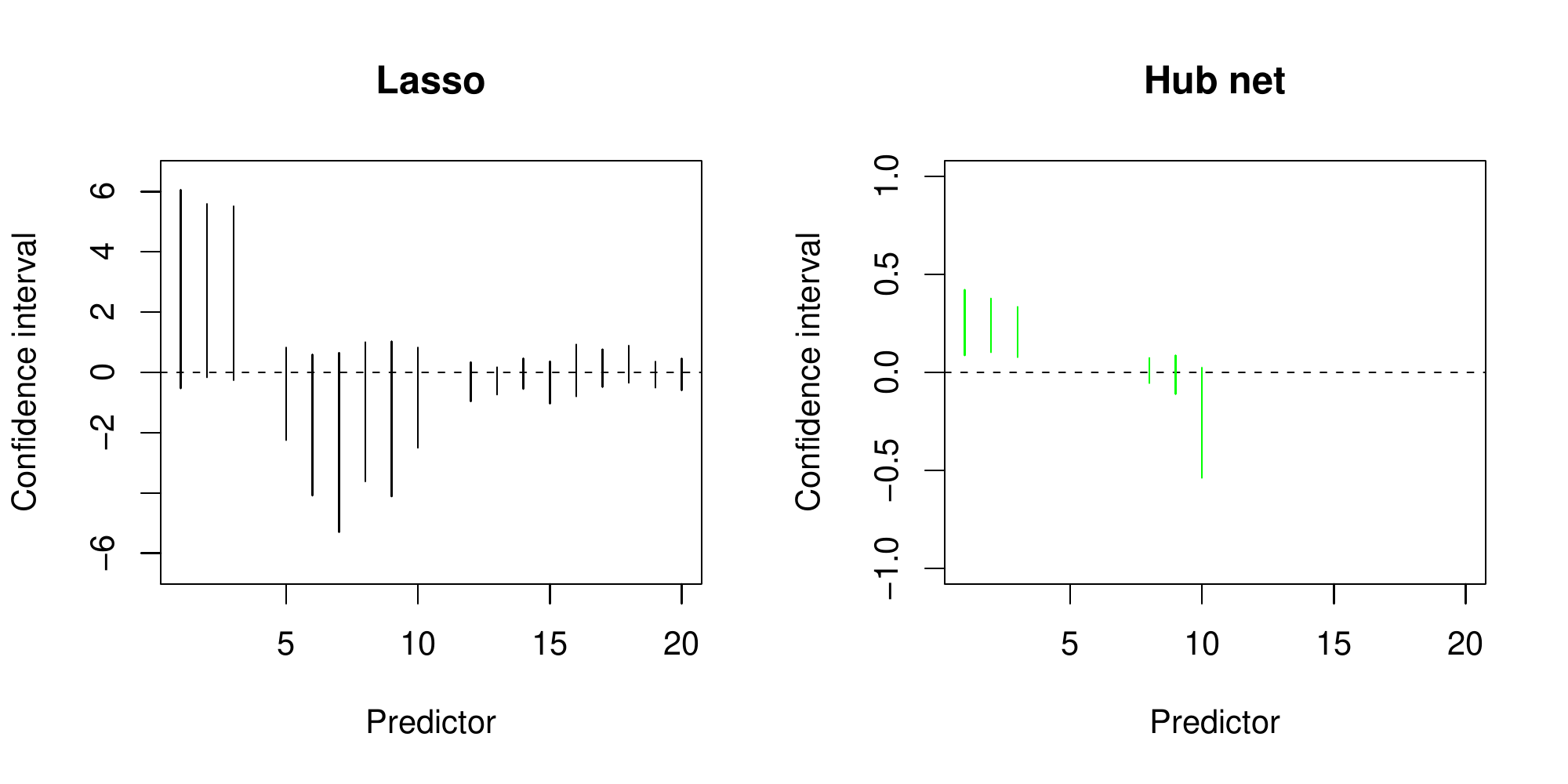}
\caption[figinf]{\em 90\% post-selection confidence intervals for a realization from the setting of Figure \ref{fig:fig1},
for lasso (left panel) and hubNet (right panel). Note the different vertical scales in the two plots.
 }
\label{fig:figinf}
\end{center}
\end{figure}

Since the construction of weights in the  hubNet procedure is unsupervised, we can apply recently developed
post-selection inference tools for the lasso. In particular, \citet{lee2016exact} construct p-values and confidence intervals 
for the lasso that have exact type I error control and coverage, conditional on the active set of predictors chosen.
We  can apply these methods to the output of  hubNet, since the estimation is just a lasso with weights.
Figure  \ref{fig:figinf} shows the 90\% post-selection confidence intervals for a realization from the setting of Figure \ref{fig:fig1},
for lasso (left panel) and hubNet (right panel). For the lasso, we see there are
no coefficients whose intervals are away from zero, and the intervals are very wide.
The hubNet intervals are much shorter, and correctly detect the non-zero coefficients (first three predictors).

\section{Recovery of hub nodes and speed comparisons}
\label{sec:hubmethodcomparison}
In this section, we compare the edge-out method with the {\tt hglasso} method of
\cite{tanetal} in terms of computational speed and recovery of the underlying
structure. We generate $\bX$ according to three settings:

\begin{enumerate}
\item For a core set $S$ of size $s$, let
$\bA \in \{0,1\}^{p \times p}$ have all diagonal
entries 1, all entries in row $i$ and column $i$ equal to 1 for all $i \in S$,
and remaining entries 0. Define
\[\bE=\begin{cases}
0 &  \bA_{ij} = 0\\
\operatorname{Unif}([-0.15, -0.015]\cup [0.015,0.15])& \text{otherwise,}
\end{cases}\]
$\bar{\bE} = \frac{1}{2}(\bE+\bE^T)$, and $\bSigma^{-1}=\bar{\bE}
+(0.2-\lambda_{\min}(\bar{\bE}))\Id$,
and generate the rows of $\bX$ from $N(0,\bSigma)$. \\

\item For two predictor sets $S_1$ and $S_2$ of sizes $s/2$, let
\[\bA=\begin{pmatrix} \bA_1 & 0\\ 0 & \bA_2 \end{pmatrix}\]
with $\bA_1,\bA_2$ generated as above with core sets $S_1,S_2$.
Construct $\bX$ from $\bA$ in the same way as above.

\item For a core set $S$ of size $s$, generate $\bGamma \in \real^{s \times
(p-s)}$ with i.i.d.\ entries distributed as $N(0,4)$ truncated above and below
at $\pm 2$. Then generate each row $\bX_{i,.}$ of $\bX$ such that
$\bX_{ij}\sim N(0,1)$ for $j \in S$ and $\bX_{ij}=\bX_{i,S}
\bGamma_{.,j}+\epsilon_{ij}$ for $j \notin S$ and $\epsilon_{ij} \sim N(0,1)$.
\end{enumerate}
In each setting, we re-standardize the predictors to have variance 1.

\begin{figure}[H]
\begin{center}
\includegraphics[width=0.8\textwidth]{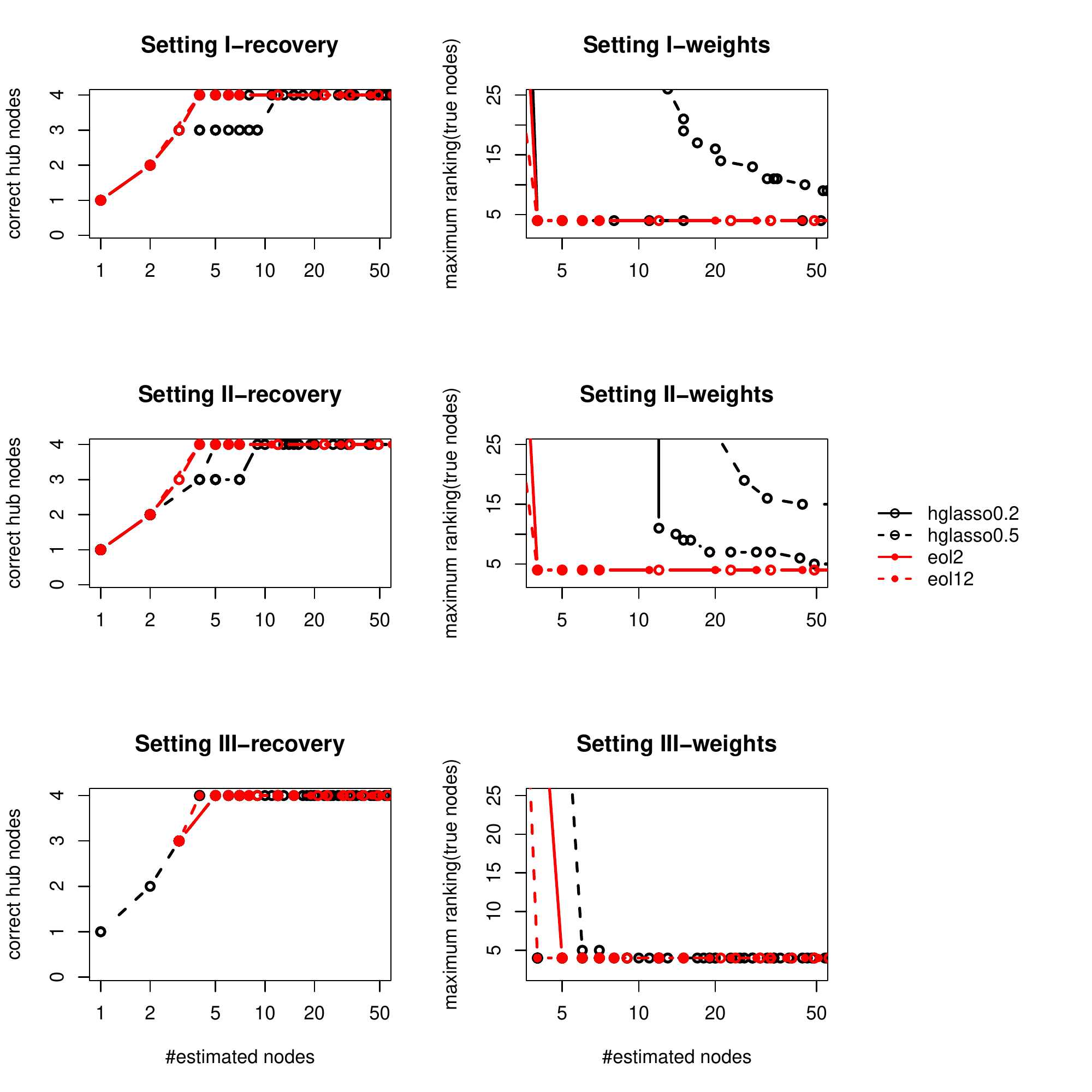}
\caption{\em Recovery results and weights ranking}
\label{fig:edgeoutglassorecovery}
\end{center}
\end{figure}

In Figure \ref{fig:edgeoutglassorecovery}, we set $(n,p,s)=(100,200,4)$
and compare edge-out and hglasso by the number of correctly identified hub
nodes as well as their corresponding absolute row sums in the estimated matrix.
(This matrix is $\hat{\bB}_{eo}$ for edge-out and $\hat{\bV}^T$ in the
hglasso decomposition $\bSigma^{-1}=\bZ+\bV+\bV^T$ where $\bZ$ is sparse and
$\bV^T$ has few non-zero rows.) Edge-out was applied with only the $\ell_2$
penalty (eol2) or with $\gamma=0.5$ (eol12), and hglasso with $\lambda_1=1000$
and $\lambda_2=0.2$ or 0.5. The left column of the figure tracks the number
of correctly identified hubs as the main tuning parameter ($\theta$ for edge-out
and $\lambda_3$ for hglasso) varies, while the right
column tracks the maximum rank of any hub node when all nodes are ranked in
decreasing order of their absolute row sums. (A maximum rank of 4 indicates
that all four hub nodes have larger absolute row sums than all remaining nodes.)
Both variants of edge-out perform well in all three settings; hglasso performs
well in settings 1 and 3 for $\lambda_2=0.2$ but not for setting 2 under the
tested tuning parameters.

\begin{figure}[H]
 \begin{center}
 \includegraphics[width=.7\textwidth]{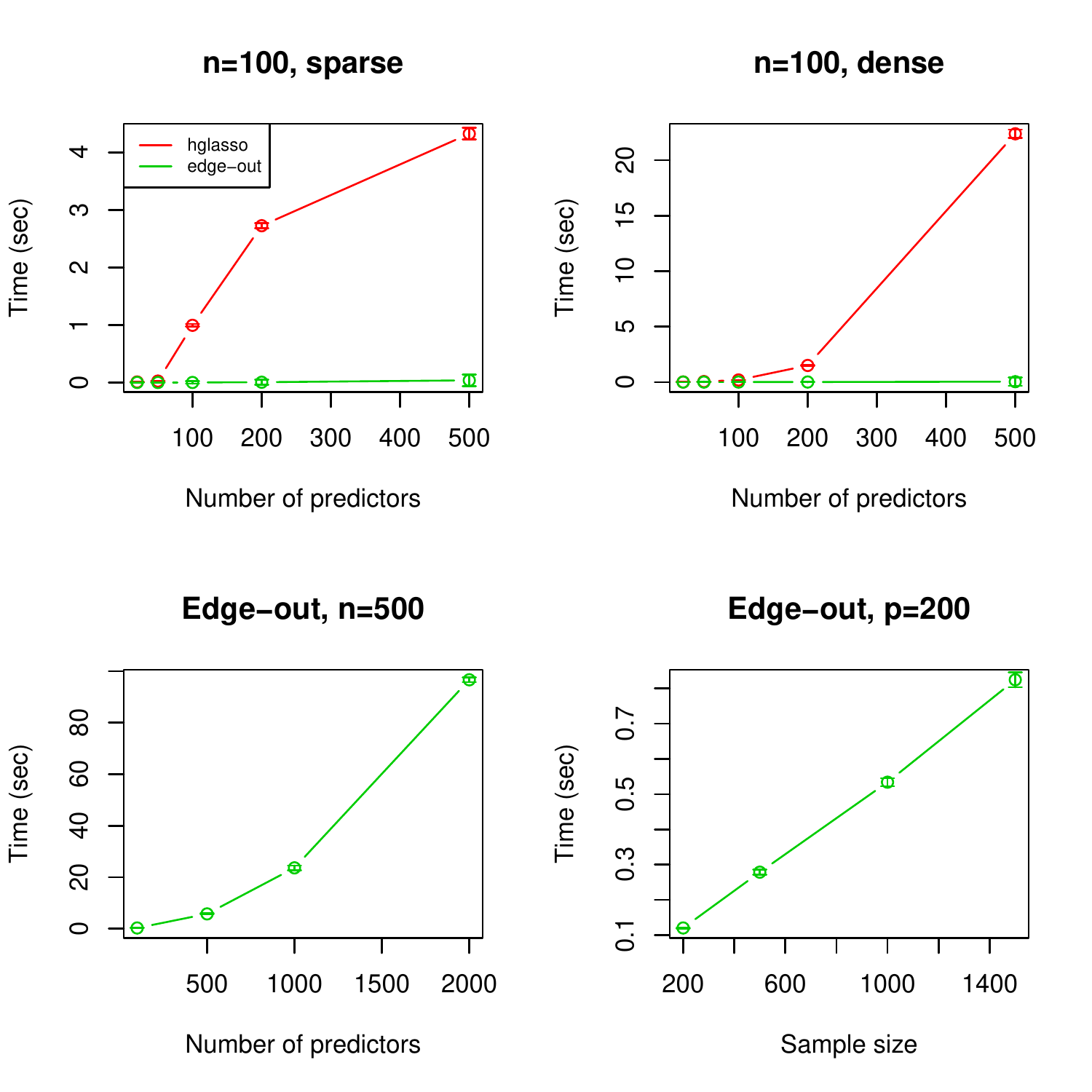}
 \end{center}
 \caption[fig:figspeed]{\em Speed comparisons. In the top row we compare the
computation times for the {\tt hglasso} and {\tt edge-out }algorithms, as the number of predictors increases,
 for sparse and dense problems. The bottom row examines just {\tt edge-out}, with $n$ or $p$ fixed, for larger problems. We were not able to run {\tt hglasso} in these latter settings.}
 \label{fig:figspeed}
 \end{figure}

Figure \ref{fig:figspeed} compares the speed of these two methods, with one of
$n,p$ fixed while the other grows. We see that the edge-out algorithm is much
faster and appears to scale quadratically in $p$ and linearly in $n$.

 \section{Discussion}
 We have proposed a new procedure, hubNet, that is applicable to many supervised learning problems.
The procedure estimates ``hub weights'' from the matrix of predictor values and then
 uses these weights in a supervised learning method such as the lasso or random
forest.

HubNet provides a way of utilizing structural information in the predictors,
and it can yield more accurate prediction and support recovery in certain situations
known to be hard if we neglect such knowledge. Since the estimation of
weights is done in an unsupervised
manner, standard cross-validation can be applied in the weighted fitting step. 
We observe in practice that this new procedure can sometimes yield lower
prediction 
 error than the unweighted approach, or give similar prediction
error using fewer features. Moreover, the estimation of the hub structure can
also be useful for interpretation.

 Further work is needed in making the edge-out algorithm for hub estimation more
efficient, so that it can be applied to very large datasets. 
 \medskip
 
 {\bf Acknowledgments}

Zhou Fan was supported supported by a Hertz Foundation Fellowship and an NDSEG
Fellowship (DoD AFOSR 32 CFR 168a).
Robert Tibshirani was supported by NIH grant 5R01 EB001988-16 and NSF grant DMS1208164.
 
\appendix
\section{Optimization for the edge-out model}\label{app:optimization}
We consider the objective function (\ref{eqn:edgeout}). The diagonal elements
of $\bB$ are fixed at zero. Let $\bX_{.,i}$ and $\bX_{.,-i}$ denote the $i$th
column of $\bX$ and $\bX$ with $i$th column removed, and let $\bB_{-i,-i}$
denote $\bB$ with $i$th row and column both removed.
Let $S(x,t)={\rm sign}(x)(|x|-t)_+$ be the soft-thresholding operator.

We use the following blockwise coordinate descent algorithm similar to that
of \cite{PZetal2008}:
\noindent
\begin{samepage}
\begin{enumerate}
\item Initialize $\bB=0$.
\item Iterate over $i \in \{1,2,\ldots,p\}$ until convergence:
\begin{enumerate}
\item Compute the $1 \times (p-1)$ vector
$\mathbf{r}_{i,-i}=\bX_{.,i}^T(\bX_{.,-i}-\bX_{.,-i}\bB_{-i,-i})$.
\item Compute the elementwise soft-thresholded vector
$\beta_{i,-i}=S(\mathbf{r}_{i,-i},\theta\gamma)$.
\item Update the $i$th row of $\bB$:
\[
\bB_{i,-i}=\begin{cases} 0 & \|\beta_{i,-i}\|_2\|\bX_{.,i}\|^2_2 \leq \theta(1-\gamma)
\sqrt{p-1}\\
(1-\frac{\theta(1-\gamma)\sqrt{p-1}}{\|\beta_{i.-i}\|_2\|\bX_{.,i}\|^2_2})\beta_{i,-i} &
\|\beta_{i,-i}\|_2\|\bX_{.,i}\|^2_2 > \theta(1-\gamma)\sqrt{p-1}
\end{cases}
\]
\end{enumerate}
\end{enumerate}
It can be shown that, fixing all entries of $\bB$ not in row $i$, the above
update expression exactly minimizes the objective over $\bB_{i,-i}$. Then
this procedure is a blockwise coordinate descent algorithm, applied to an
objective whose non-differentiable component is separable across
blocks, and hence converges to the solution.

\end{samepage}

\section{Proof of Theorems \ref{thm:thmA} and
\ref{thm:thmB}}\label{app:proof}
Denote by $\bX_S$ and $\bX_{S^C}$ the submatrices of $\bX$
consisting of predictors in $S$ and $S^C$, and define
\[\hat{\bSigma}_{SS}:=\frac{1}{n} \bX_S^T\bX_S,\;\;\;\;
\hat{\bSigma}_{S^CS}:=\frac{1}{n} \bX_{S^C}^T\bX_S,\;\;\;\;
\bW:=\bX_{S^C}-\bX_S\bGamma.\]
Note that by (\ref{eqn:model}), $\bW$ is independent of $\bX_S$ with independent
Gaussian entries of variance at most 1. The following lemma collects
probabilistic statements involving $\bX_S$ and $\bW$; its proof is deferred to
Appendix \ref{app:proofdetails}.
\begin{lemma}\label{lemma:probabilitybounds}
Suppose $n,p \to \infty$, $1 \leq s \leq p$, and $s \ll n$.
If $\lambda_{\min}(\bSigma_{SS}) \geq C_{\min}$ for a constant $C_{\min}>0$,
then each of the following statements holds with probability approaching 1:
\begin{align}
\max_{j=1}^p \|\bX_{.,j}\|^2& \leq 2n+6\log p\label{eq:maxXXfull}\\
\max_{j=1}^s \|\bX_{.,j}\|^2& \leq 2n\label{eq:maxXX}\\
\|\hat{\bSigma}^{-1}_{SS}\|_2& \leq 2C_{\min}^{-1}\label{eq:S2}\\
\|\hat{\bSigma}^{-1}_{SS}\|_\infty& \leq \|\bSigma_{SS}^{-1}\|_\infty
+3(s+\sqrt{s}\log n)/(C_{\min}\sqrt{n})\label{eq:Sinfty}\\
\|\hat{\bSigma}_{SS}^{-1}\bX_S^T\bW\|_{\infty,2}&\leq\sqrt{4np/C_{\min}}
\label{eq:SXWinfty2}\\
\|\bW^T\bX_S\hat{\bSigma}_{SS}^{-1}\|_{\infty,2}&\leq\sqrt{4n(s+3\log p)/
C_{\min}}\label{eq:WXSinfty2}\\
\|\bW^T(\Id_{s \times s}-\tfrac{1}{n}\bX_S\hat{\bSigma}_{SS}^{-1}\bX_S^T)
\bW\|_{\infty,2}&\leq 2n+\sqrt{3np}+\sqrt{6p\log p}.\label{eq:WPWinfty2}
\end{align}
\end{lemma}

\subsection*{Proof of Theorem \ref{thm:thmA}}
Our proof draws upon a similar analysis of support recovery in the multivariate
regression setting by \cite{obozinski2011support}.
Let us introduce $\theta_n=\theta\sqrt{p-1}/n$ and write the edge-out
estimate (in the case $\gamma=0$) as
\begin{equation}\label{eq:edgeoutrescaled}
\hat{\bB}_{eo}=\argmin_{\bB \in \real^{p \times p}:\,\bB_{ii}=0\,\forall i}
\frac{1}{2n}\|\bX-\bX \bB\|^2_F+\theta_n \sum_{i=1}^p \|\bB_{i,.}\|_2.
\end{equation}
Consider the restricted problem over $\bB \in \real^{s \times p}$ where each
predictor is regressed only on $\bX_S$:
\begin{align}\label{eq:restricted}
\hat{\bB}_{\text{restricted}}&=\argmin_{\bB \in \real^{s \times p}:\,
\bB_{ii}=0\,\forall i} \frac{1}{2n}\|\bX-\bX_S\bB\|^2_F
+\theta_n\sum_{i \in S} \|\bB_{i,.}\|_2.
\end{align}
The subgradient conditions for optimality of $\hat{\bB}_{eo}$ and
$\hat{\bB}_{\text{restricted}}$ imply the following sufficient condition for
recovery of $S$, whose proof we defer to Appendix \ref{app:proofdetails}:
\begin{lemma}\label{lemma:fullsubgradient}
If $\bX_S^T\bX_S$ is invertible, then the solution
$\hat{\bB}:=\hat{\bB}_{\text{restricted}}$ to (\ref{eq:restricted}) is unique.
If furthermore this solution satisfies
\begin{align}
\max_{j \in S^c} \frac{1}{n}\|\bX_{\cdot,j}^T(\bX-\bX_S\hat{\bB})\|_2
&<\theta_n,\label{eq:condition1}\\
\min_{i \in S}\|\hat{\bB}_{i,.}\|_2 &>0,\label{eq:condition2}
\end{align}
then the solution $\hat{\bB}_{eo}$ to (\ref{eq:edgeoutrescaled}) is unique,
with the first $s$ rows non-zero and equal to $\hat{\bB}$ and
remaining rows equal to 0.
\end{lemma}

Through the remainder of this appendix, let
$\hat{\bB}:=\hat{\bB}_{\text{restricted}} \in \real^{s \times p}$ be the
solution to the restricted problem (\ref{eq:restricted}). 
As $s \ll n$ and $\bSigma_{SS}$ is non-singular, $\bX_S^T\bX_S$ is invertible
with probability 1. Hence, to prove Theorem \ref{thm:thmA},
it suffices to show that (\ref{eq:condition1})
and (\ref{eq:condition2}) hold with high probability. Define
\begin{align*}
\bU &:= \begin{pmatrix} \Id_{s \times s} &
\frac{1}{n}\hat{\bSigma}_{SS}^{-1}\bX_S^T\bW \end{pmatrix} \in \real^{s \times
p},\\
\bB^* &:= \begin{pmatrix} \mathbf{0}_{s \times s} & \bGamma \end{pmatrix} \in
\real^{s \times p},\\
\hat{\bD}&:=\operatorname{diag}\left(\|\hat{\bB}_{1,.}\|^{-1}_2,...,
\|\hat{\bB}_{s,.}\|^{-1}_2\right) \in \real^{s \times s},\\
\bDelta &\in \real^{s \times p},\;\; \bDelta_{ij}:=
\begin{cases} \bX^T_{.,j}(\bX_{.,j} - \bX_S\hat{\bB}_{.,j}) & i=j\\
0 & \text{otherwise},\end{cases}\\
\mathcal{Z}&:=\left\{\bZ \in [-1,1]^{s \times p}:
\begin{array}{ll}
\bZ_{i,.}=\hat{\bD}_{i,i}\hat{\bB}_{i,.}
&\mbox{ if }\|\hat{\bB}_{i,.}\|_2 > 0\\
\bZ_{i,i}=0 \text{ and } \|\bZ_{i,.}\|_2 \leq 1&\mbox{ if }
\|\hat{\bB}_{i,.}\|_2=0
\end{array}\right\}
\end{align*}
The subgradient condition for optimality of $\hat{\bB}$ for
(\ref{eq:restricted}) implies the following, whose proof we also defer to
Appendix \ref{app:proofdetails}.
\begin{lemma}\label{lemma:restrictedsubgradient}
There exists $\bZ \in \mathcal{Z}$ such that
\[\hat{\bB}-\bB^*=\bU-\theta_n \hat{\bSigma}_{SS}^{-1}\bZ
-\tfrac{1}{n}\hat{\bSigma}_{SS}^{-1}\bDelta.\]
\end{lemma}

Using these lemmas, we now verify conditions (\ref{eq:condition1}) and
(\ref{eq:condition2}):
\begin{lemma}\label{lemma:condition2}
Suppose Assumptions \ref{assump:Cmin}, \ref{assump:s}, and \ref{assump:Gmin}
hold, and $\theta_n$ satisfies (\ref{eq:lambdachoice}).
Then with probability approaching 1, (\ref{eq:condition2}) holds and
\[\|\hat{\bB}-\bB^*\|_{\infty,2} \leq 2\theta_n\|\bSigma_{SS}^{-1}\|_\infty.\]
\end{lemma}
{\bf Proof:}

By Lemma \ref{lemma:restrictedsubgradient}, for some $\bZ \in \mathcal{Z}$,
 \[\|\hat{\bB}-\bB^*\|_{\infty,2} \leq \|\bU\|_{\infty,2}+
\theta_n\|\hat{\bSigma}_{SS}^{-1}\bZ\|_{\infty,2}
+\tfrac{1}{n}\|\hat{\bSigma}_{SS}^{-1}\bDelta\|_{\infty,2}.\]
For the first term, (\ref{eq:SXWinfty2}) and the definition of $\bU$
imply, with probability approaching 1,
\[\|\bU\|_{\infty,2} \leq 1+\sqrt{4p/(C_{\min}n)}.\]
For the second term, (\ref{eq:Sinfty}) and the observation
$\|\bZ\|_{\infty,2} \leq 1$ imply, with probability approaching 1,
\[\|\hat{\bSigma}_{SS}^{-1}\bZ\|_{\infty,2}
\leq \|\hat{\bSigma}_{SS}^{-1}\|_\infty\|\bZ\|_{\infty,2}
\leq \|\hat{\bSigma}_{SS}^{-1}\|_\infty \leq
\|\bSigma_{SS}^{-1}\|+3(s+\sqrt{s}\log n)/(C_{\min}\sqrt{n}).\]
For the third term, note that for all $j=1,\ldots,p$,
\begin{equation}\label{eq:Deltajj}
|\bDelta_{jj}| \leq \|\bX_{.,j}\|^2,
\end{equation}
for otherwise
\[\|\bX_{.,j}-\bX_S\hat{\bB}_{.,j}\|_2^2-\|\bX_{.,j}\|^2
=(2\bX_{.,j}-\bX_S\hat{\bB}_{.,j})^T(-\bX_S\hat{\bB}_{.,j})>0,\]
implying that the objective (\ref{eq:restricted}) would decrease upon setting
$\hat{\bB}_{.,j}=0$ and contradicting optimality of $\hat{\bB}$. Then, as
$\bDelta$ is diagonal, (\ref{eq:maxXX}) and (\ref{eq:S2}) imply,
with probability approaching 1,
\[\|\hat{\bSigma}_{SS}^{-1}\bDelta\|_{\infty,2}
\leq \|\hat{\bSigma}_{SS}^{-1}\|_{\infty,2}\max_{j=1}^s |\bDelta_{jj}|
\leq \|\hat{\bSigma}_{SS}^{-1}\|_2\max_{j=1}^s \|\bX_{.,j}\|_2^2
\leq 4n/C_{\min}.\]
Noting that $\|\bSigma_{SS}^{-1}\|_\infty \geq \|\bSigma_{SS}^{-1}\|_2=
1/\lambda_{\min}(\bSigma_{SS}) \geq 1$ by our normalization $\bSigma_{jj}=1$ for
all $j$, we have under the given assumptions
\[\max(1,\sqrt{p/n},\theta_n s/\sqrt{n},\theta_n \sqrt{s/n}\log n,
\ll \theta_n \|\bSigma_{SS}^{-1}\|_\infty \ll \Gamma_{\min}.\]
Then with probability approaching 1, $\|\hat{\bB}-\bB^*\|_{\infty,2} \leq
2\theta_n\|\bSigma_{SS}^{-1}\|_\infty$ and
\[\min_i \|\hat{\bB}_{i,.}\|_2 \geq \min_i
\|\bB^*_{i,.}\|_2-2\theta_n\|\bSigma_{SS}^{-1}\|_\infty
=\Gamma_{\min}-2\theta_n\|\bSigma_{SS}^{-1}\|_\infty>0.\]
$\blacksquare$
\begin{lemma}\label{lemma:condition1}
Suppose Assumptions \ref{assump:Cmin}, \ref{assump:irrepresentable},
\ref{assump:s}, and \ref{assump:Gmin} hold, and $\theta_n$
satisfies (\ref{eq:lambdachoice}). Then (\ref{eq:condition1}) holds
with probability approaching 1.
\end{lemma}
{\bf Proof:}
By Lemma \ref{lemma:condition2}, it suffices to consider the event where
$\|\hat{\bB}_{i,.}\|_2>0$ for all $i \in S$, and hence $\bZ=\hat{\bD}\hat{\bB}$
in Lemma \ref{lemma:restrictedsubgradient}.
On this event, writing $\bX=(\bX_S,\bX_S\bGamma+\bW)=(\bX_S,\bW)+\bX_S\bB^*$
and applying Lemma \ref{lemma:restrictedsubgradient},
\begin{align}
\frac{1}{n}\|\bX^T_{S^C}&(\bX - \bX_S\hat{\bB})\|_{\infty,2}
=\frac{1}{n}\|\bX^T_{S^C}(\bX_S,\bW)+\bX_{S^C}^T\bX_S(\bB^*-\hat{\bB})
\|_{\infty,2}\nonumber\\
&\leq \frac{1}{n}\|\bX^T_{S^C}(\bX_S,\bW)-\bX_{S^C}^T\bX_S\bU\|_{\infty,2}
+\theta_n\|\hat{\bSigma}_{S^CS}\hat{\bSigma}_{SS}^{-1}
\hat{\bD}\hat{\bB}\|_{\infty,2}+\frac{1}{n}\|\hat{\bSigma}_{S^CS}\hat{\bSigma}_{SS}^{-1}\bDelta\|_{\infty,2}\label{eq:condition1decomp}.
\end{align}

For the first term of (\ref{eq:condition1decomp}), recalling the definition of
$\bU$, noting that $\bX_S^T(\Id-\frac{1}{n}\bX_S\hat{\bSigma}_{SS}^{-1}\bX_S^T)
=0$, and applying (\ref{eq:WPWinfty2}), with probability approaching 1,
\begin{align*}
\|\bX^T_{S^C}(\bX_S,\bW)&-\bX_{S^C}^T\bX_S\bU\|_{\infty,2}
=\|\bX_{S^C}^T(\Id-\tfrac{1}{n}\bX_S\hat{\bSigma}_{SS}^{-1}\bX_S^T)
\bW\|_{\infty,2}\\
&=\|\bW^T(\Id-\tfrac{1}{n}\bX_S\hat{\bSigma}_{SS}^{-1}\bX_S^T)\bW\|_{\infty,2}
\leq 2n+\sqrt{3np}+\sqrt{6p\log p} \ll n\theta_n.
\end{align*}
For the third term of (\ref{eq:condition1decomp}), applying
(\ref{eq:Deltajj}), (\ref{eq:Gammacolbound}),
(\ref{eq:maxXX}), and (\ref{eq:WXSinfty2}), with probability approaching 1,
\begin{align*}
\|\hat{\bSigma}_{S^CS}\hat{\bSigma}_{SS}^{-1}\bDelta\|_{\infty,2}
&\leq \|\hat{\bSigma}_{S^CS}\hat{\bSigma}_{SS}^{-1}\|_{\infty,2}\max_{j=1}^s
|\bDelta_{jj}|
=\frac{1}{n}\|(\bX_S\bGamma+\bW)^T\bX_S\hat{\bSigma}_{SS}^{-1}\|_{\infty,2}
\max_{j=1}^s |\bDelta_{jj}|\\
&\leq \left(\|\bGamma^T\|_{\infty,2}+\frac{1}{n}
\|\bW^T\bX_S\hat{\bSigma}_{SS}^{-1}\|_{\infty,2}\right)
\max_{j=1}^s \|\bX_{.,j}\|_2^2
\leq \frac{2n}{\sqrt{C_{\min}}}+\sqrt{\frac{16n(s+3\log
p)}{C_{\min}}} \ll n\theta_n.
\end{align*}

It remains to bound the second term of (\ref{eq:condition1decomp}). Let
$\bD$ be as in Assumption \ref{assump:irrepresentable} and write
\begin{align*}
\hat{\bSigma}_{S^CS}\hat{\bSigma}_{SS}^{-1}\hat{\bD}\hat{\bB}
&=\bGamma^T\bD\bB^*+\bGamma^T\bD(\hat{\bB}-\bB^*)
+\bGamma^T(\hat{\bD}-\bD)\hat{\bB}+(\hat{\bSigma}_{S^CS}\hat{\bSigma}_{SS}^{-1}
-\bGamma^T)\hat{\bD}\hat{\bB}\\
&=:\mathbf{I}+\mathbf{II}+\mathbf{III}+\mathbf{IV}.
\end{align*}
By Assumption \ref{assump:irrepresentable} and the definition of $\bB^*$,
\[\|\mathbf{I}\|_{\infty,2}=\|\bGamma^T\bD\bGamma\|_{\infty,2} \leq 1-\delta.\]
By Lemma \ref{lemma:condition2}, with probability approaching 1,
\[\|\mathbf{II}\|_{\infty,2}
\leq \|\bGamma^T\|_\infty\|\bD(\hat{\bB}-\bB^*)\|_{\infty,2}
\leq \|\bGamma^T\|_\infty\Gamma_{\min}^{-1}\|\hat{\bB}-\bB^*\|_{\infty,2}
\leq 2\|\bGamma^T\|_\infty\Gamma_{\min}^{-1}\theta_n
\|\bSigma_{SS}^{-1}\|_\infty \ll 1.\]
$\mathbf{III}$ satisfies the same bound, as
\[\|\mathbf{III}\|_{\infty,2}\leq
\|\bGamma^T\|_\infty \|(\hat{\bD}-\bD)\hat{\bB}\|_{\infty,2}
=\|\bGamma^T\|_\infty \max_{i \in S}
\frac{|\|\bB^*_{i,.}\|_2-\|\hat{\bB}_{i,.}\|_2|}{\|\bB^*_{i,.}\|_2}
\leq \|\bGamma^T\|_\infty \|\bD(\hat{\bB}-\bB^*)\|_{\infty,2}.\]
Finally, using $\bX_{S^C}=\bX_S\bGamma+\bW$ and applying (\ref{eq:WXSinfty2}),
with probability approaching 1,
\begin{align*}
\|\mathbf{IV}\|_{\infty,2}&=\|(\tfrac{1}{n}\bX_{S^C}^T\bX_S
\hat{\bSigma}_{SS}^{-1}-\bGamma^T)\hat{\bD}{\hat{\bB}}\|_{\infty,2}
=\frac{1}{n}\|\bW^T\bX_S\hat{\bSigma}_{SS}^{-1}
\hat{\bD}\hat{\bB}\|_{\infty,2}\\
&\leq \frac{1}{n}\|\bW^T\bX_S\hat{\bSigma}_{SS}^{-1}\|_\infty
\|\hat{\bD}\hat{\bB}\|_{\infty,2}
\leq \frac{\sqrt{s}}{n}\|\bW^T\bX_S\hat{\bSigma}_{SS}^{-1}\|_{\infty,2}
\leq \sqrt{\frac{4s(s+3\log p)}{C_{\min}n}} \ll 1.
\end{align*}
Combining the above yields
$\|\hat{\bSigma}_{S^CS}\hat{\bSigma}_{SS}^{-1}\hat{\bD}\hat{\bB}\|_{\infty,2}
\leq 1-\delta/2$
with probability approaching 1, which together with (\ref{eq:condition1decomp})
implies (\ref{eq:condition1}).

$\blacksquare$

Theorem \ref{thm:thmA} follows from Lemmas \ref{lemma:fullsubgradient},
\ref{lemma:condition2}, and \ref{lemma:condition1}.

\subsection*{Proof of Theorem~\ref{thm:thmB}} 
We verify the conditions of Lemma 8.2 of \citet{zhou2009adaptive} under the
given assumptions and in our asymptotic setting with random design.
By (\ref{eq:maxXXfull}) and (\ref{eq:S2}), with probability approaching 1,
\begin{equation}\label{eq:thmBcond1}
\max_{j \in S^C} \frac{\|\bX_{.,j}\|_2}{\sqrt{n}} \leq \sqrt{2+\frac{6\log
p}{n}},\;\;\;\; \lambda_{\min}(\hat{\bSigma}_{SS}) \geq \frac{C_{\min}}{2}.
\end{equation}
It remains to verify the weighted incoherency condition (8.4a) of
\citet{zhou2009adaptive}.
Define $\bD_{w,S}=\operatorname{diag}(w_1,\ldots,w_s) \in \real^{s \times s}$
and $\bD_{w,S^C}^{-1}=\operatorname{diag}(w_{s+1}^{-1},\ldots,w_p^{-1}) \in
\real^{(s-p) \times (s-p)}$ where $w_k^{-1}=0$ if $w_k=\infty$. Then
\[\|\bD_{w,S^C}^{-1}\bX_{S^C}^T\bX_S(\bX_S^T\bX_S)^{-1}\bD_{w,S}\|_\infty
\leq \frac{w_{\max}(S)}{nw_{\min}(S^C)}
\|\bX_{S^C}^T\bX_S\hat{\bSigma}_{SS}^{-1}\|_\infty
\leq \frac{\rho}{n}\|\bX_{S^C}^T\bX_S\hat{\bSigma}_{SS}^{-1}\|_\infty.\]
Writing $\bX_{S^C}=\bX_S\bGamma+\bW$ and applying (\ref{eq:Gammacolbound})
and (\ref{eq:WXSinfty2}), with probability approaching 1,
\begin{align*}
\frac{1}{n}\|\bX_{S^C}^T\bX_S\hat{\bSigma}_{SS}^{-1}\|_\infty
&\leq \frac{\sqrt{s}}{n}\|\bX_{S^C}^T\bX_S\hat{\bSigma}_{SS}^{-1}\|_{\infty,2}
\leq \sqrt{s}\|\bGamma^T\|_{\infty,2}
+\frac{\sqrt{s}}{n}\|\bW^T\bX_S\bSigma_{SS}^{-1}\|_{\infty,2}\\
&\leq \sqrt{\frac{s}{C_{\min}}}+\sqrt{\frac{4s(s+3\log p)}{nC_{\min}}}
\leq \sqrt{\frac{s}{C_{\min}}}\left(1+\sqrt{\frac{12\log p}{n}}+o(1)\right).
\end{align*}
Hence under Assumption \ref{assump:wgap}, with probability approaching 1,
\begin{equation}\label{eq:thmBcond2}
\|\bD_{w,S^C}^{-1}\bX_{S^C}^T\bX_S(\bX_S^T\bX_S)^{-1}\bD_{w,S}\|_\infty
\leq 1-\eta-o(1) \leq 1-\eta/2.
\end{equation}
Conditional on $\bX$, on the event where (\ref{eq:thmBcond1}) and
(\ref{eq:thmBcond2}) hold, our conclusion follows from
Lemma 8.2 of \citet{zhou2009adaptive}. Then the conclusion also follows
unconditionally.

\section{Proofs of supporting lemmas}\label{app:proofdetails}
In this appendix, we prove Lemmas \ref{lemma:probabilitybounds},
\ref{lemma:fullsubgradient}, and \ref{lemma:restrictedsubgradient}.

\subsection*{Proof of Lemma \ref{lemma:probabilitybounds}}
Our normalization $\bSigma_{jj}=1$ implies $\|\bX_{.,j}\|_2^2 \sim \chi^2_n$ for
each $j=1,\ldots,p$. We use the chi-squared tail bound
\begin{equation}\label{eq:chisqtail}
P[\chi_n^2>n+2\sqrt{nt}+2t] \leq \exp(-t)
\end{equation}
for all $t>0$, from Lemma 1 of \citet{laurentmassart}. Then
\[P[\|\bX_{.,j}\|_2^2>2n+6\log p] \leq
P[\|\bX_{.,j}\|_2^2>n+2\sqrt{2n\log p}+4\log p] \leq \exp(-2\log p),\]
and a union bound over $j=1,\ldots,p$ yields (\ref{eq:maxXXfull}).
Also, $P[\|\bX_{.,j}\|_2^2>2n] \leq \exp(-n/8)$, and as $s \ll n$,
a union bound over $j=1,\ldots,s$ yields (\ref{eq:maxXX}).
For (\ref{eq:S2}) and (\ref{eq:Sinfty}),
\[\|\hat{\bSigma}_{SS}^{-1}-\bSigma_{SS}^{-1}\|_2
\leq \|\bSigma_{SS}^{-1/2}\|_2\|\bSigma_{SS}^{1/2}\hat{\bSigma}_{SS}^{-1}
\bSigma_{SS}^{1/2}-\Id\|_2\|\bSigma_{SS}^{-1/2}\|_2
\leq C_{\min}^{-1}\|\tilde{\bSigma}_{SS}^{-1}-\Id\|_2\]
where $\tilde{\bSigma}_{SS}\overset{L}{=}n^{-1}\bZ^T\bZ$ for
$\bZ \in \real^{n \times s}$ having i.i.d.\ standard Gaussian entries.
Corollary 5.35 of \citet{vershynin} implies
\[\left(1-\frac{\sqrt{s}+\log n}{\sqrt{n}}\right)^2 \leq
\lambda_{\min}(\tilde{\bSigma}_{SS}) \leq
\lambda_{\max}(\tilde{\bSigma}_{SS}) \leq \left(1+\frac{\sqrt{s}+\log
n}{\sqrt{n}}\right)^2\]
with probability approaching 1. As $s \ll n$, this implies for any $\delta>0$,
with probability approaching 1
\[\|\tilde{\bSigma}_{SS}^{-1}-\Id\|_2 \leq (2+\delta)\left(\frac{\sqrt{s}+\log
n}{\sqrt{n}}\right).\]
Then (\ref{eq:S2}) follows from $\|\hat{\bSigma}_{SS}^{-1}\|_2 \leq
\|\hat{\bSigma}_{SS}^{-1}-\bSigma_{SS}^{-1}\|_2+\|\bSigma_{SS}^{-1}\|_2
\leq 2C_{\min}^{-1}$, and (\ref{eq:Sinfty}) from
\[\|\hat{\bSigma}_{SS}^{-1}\|_\infty
\leq \|\hat{\bSigma}_{SS}^{-1}-\bSigma_{SS}^{-1}\|_\infty
+\|\bSigma_{SS}^{-1}\|_\infty
\leq \sqrt{s}\|\hat{\bSigma}_{SS}^{-1}-\bSigma_{SS}^{-1}\|_2
+\|\bSigma_{SS}^{-1}\|_\infty \leq \frac{3(s+\sqrt{s}\log n)}{C_{\min}\sqrt{n}}
+\|\bSigma_{SS}^{-1}\|_\infty.\]

For the remaining three statements, denote
$\bS=\operatorname{diag}(\sigma_{j+1},\ldots,\sigma_p) \in
\real^{(p-s) \times (p-s)}$, so $\bW=\bZ\bS$ where $\bZ \in \real^{n \times
(p-s)}$ is independent of $\bX_S$ with i.i.d.\ standard Gaussian entries.
Denote
$\bP=\frac{1}{\sqrt{n}}\hat{\bSigma}_{SS}^{-1/2}\bX_S^T$, so that
$\bP^T\bP$ is the projection in $\real^n$ onto the column span of $\bX_S$.
With probability 1, this column span is of rank $s$, so $\bP$ is an
orthogonal projection from $\real^n$ to $\real^s$. Applying $\sigma_j \leq 1$
for each $j$,
\[\|\hat{\bSigma}_{SS}^{-1}\bX_S^T\bW\|_{\infty,2}
=\sqrt{n}\|\hat{\bSigma}_{SS}^{-1/2}\bP\bZ\bS\|_{\infty,2}
\leq \sqrt{n}\|\hat{\bSigma}_{SS}^{-1/2}\bP\bZ\|_{\infty,2}.\]
Conditional on $\bX_S$, the columns of $\hat{\bSigma}_{SS}^{-1/2}\bP\bZ$ are
independent and distributed as $N(0,\hat{\bSigma}_{SS}^{-1})$, so each $i$th
row of $\hat{\bSigma}_{SS}^{-1/2}\bP\bZ$ consists of independent Gaussian
entries with variance $(\hat{\bSigma}_{SS}^{-1})_{ii} \leq
\|\hat{\bSigma}_{SS}^{-1}\|_2$. Then by (\ref{eq:chisqtail}),
\[P[\|(\hat{\bSigma}_{SS}^{-1/2}\bP\bZ)_{i,.}\|_2^2>2p
\|\hat{\bSigma}_{SS}^{-1}\|_2 \mid \bX_S] \leq \exp(-p/8),\]
and (\ref{eq:SXWinfty2}) follows by taking a union bound over $i=1,\ldots,s$,
recalling $s \leq p$, and applying (\ref{eq:S2}). Similarly,
$\|\bW^T\bX_S\bSigma_{SS}^{-1}\|_{\infty,2} \leq
\sqrt{n}\|\bZ^T\bP^T\bSigma_{SS}^{-1/2}\|_{\infty,2}$,
and conditional on $\bX_S$ each row of
$\bZ^T\bP^T\hat{\bSigma}_{SS}^{-1}$ is distributed as
$N(0,\hat{\bSigma}_{SS}^{-1})$. Then (\ref{eq:chisqtail}) implies
\[P[\|(\bZ^T\bP^T\hat{\bSigma}_{SS}^{-1/2})_{j,.}\|_2^2>(2s+6\log p)
\|\hat{\bSigma}_{SS}^{-1}\|_2 \mid \bX_S] \leq \exp(-2\log p),\]
and (\ref{eq:S2}) and a union bound over $j=s+1,\ldots,p$ yields
(\ref{eq:WXSinfty2}). Finally,
\[\|\bW^T(\Id-\tfrac{1}{n}\bX_S\bSigma_{SS}^{-1}\bX_S^T)\bW\|_{\infty,2}
\leq \|\bZ^T(\Id-\bP^T\bP)\bZ\|_{\infty,2},\]
and conditional on $\bX_S$, $\bZ^T(\Id-\bP^T\bP)\bZ$ is equal in law to
$\tilde{\bZ}^T\tilde{\bZ}$ where $\tilde{\bZ} \in \real^{(n-s) \times (p-s)}$
has i.i.d.\ standard Gaussian entries. Writing
$\|\tilde{\bZ}^T\tilde{\bZ}\|_{\infty,2} \leq \|\tilde{\bZ}^T\|_{\infty,2}
\|\tilde{\bZ}\|_2$, Corollary 5.35 of \citet{vershynin} implies
$\|\tilde{\bZ}\|_2 \leq \sqrt{2n}+\sqrt{p}$ with probability approaching 1,
while (\ref{eq:chisqtail}) implies
$\|\tilde{\bZ}\|_{\infty,2}^2 \leq 2n+6\log p$
with probability approaching 1. Then (\ref{eq:WPWinfty2}) follows from
combining these bounds and observing $n\log p \ll np$.

\subsection*{Proof of Lemma \ref{lemma:fullsubgradient}}
Denote by $J_{eo}(\bB)$ the objective function in (\ref{eq:edgeoutrescaled})
and by $J_{\text{restricted}}(\bB)$ the objective function in
(\ref{eq:restricted}). (The former is a function of $\bB \in \real^{p \times
p}:\bB_{ii}=0$ and the latter of $\bB \in \real^{s \times p}:\bB_{ii}=0$.)
If $\bX_S^T\bX_S$ is invertible, then $J_{\text{restricted}}$ is strictly convex
and $|J_{\text{restricted}}(\bB)| \to \infty$ as $\|\bB\|_F \to \infty$, hence
there is a unique solution $\hat{\bB}_{\text{restricted}}$ to
(\ref{eq:restricted}). Denote by $\partial J_{eo}$ and
$\partial J_{\text{restricted}}$ the
subdifferentials of $J_{eo}$ and $J_{\text{restricted}}$.
Note that $\|\bX-\bX\bB\|_F^2$ is differentiable in $\bB$
and the penalty decomposes across rows of $\bB$, hence $\partial J_{eo}(\bB)=
\mathcal{D}_1(\bB) \times \cdots \times \mathcal{D}_p(\bB)$,
where $\mathcal{D}_i(\bB)$ is the set of vectors of the form
\[-\frac{1}{n}\bX_{.,i}^T(\bX_{.,-i}-\bX\bB_{.,-i})+\theta_n
\begin{cases} \bB_{i,-i}/\|\bB_{i,-i}\|_2 & \bB_{i,-i} \neq 0 \\
\{\bZ_{i,-i}:\|\bZ_{i,-i}\|_2 \leq 1\} & \bB_{i,-i}=0 \end{cases}\]
where $\bX_{.,-i}$ and $\bB_{.,-i}$ denote $\bX$ and $\bB$ with $i$th columns
removed. Similarly, $\partial J_{\text{restricted}}(\bB)=\mathcal{D}_1(\bB)'
\times \cdots \times \mathcal{D}_s(\bB)'$ where $\mathcal{D}_i(\bB)'$ is the set
of vectors of the form
\[-\frac{1}{n}\bX_{.,i}^T(\bX_{.,-i}-\bX_S\bB_{.,-i})+\theta_n
\begin{cases} \bB_{i,-i}/\|\bB_{i,-i}\|_2 & \bB_{i,-i} \neq 0 \\
\{\bZ_{i,-i}:\|\bZ_{i,-i}\|_2 \leq 1\} & \bB_{i,-i}=0. \end{cases}\]
As $\bX\hat{\bB}_{eo}=\bX_S\hat{\bB}_{\text{restricted}}$, we have
$\mathcal{D}_i(\hat{\bB}_{eo})=\mathcal{D}_i(\hat{\bB}_{\text{restricted}})'$
for each $i \in S$. By optimality of $\hat{\bB}_{\text{restricted}}$
for (\ref{eq:restricted}),
$0 \in \partial J_{\text{restricted}}(\hat{\bB}_{\text{restricted}})$,
hence $0 \in \partial \mathcal{D}_i(\hat{\bB}_{\text{restricted}})'=
\mathcal{D}_i(\hat{\bB}_{eo})$ for each $i \in S$. On the other hand,
condition (\ref{eq:condition1}) implies $0 \in \partial
\mathcal{D}_i(\hat{\bB}_{eo})$ for each $i \in S^C$. Then
$0 \in \partial J_{eo}(\hat{\bB}_{eo})$, so $\hat{\bB}_{eo}$ solves
(\ref{eq:edgeoutrescaled}). In fact, the strict inequality in
condition (\ref{eq:condition1}) implies that 0 is in the interior of
$\mathcal{D}_i(\hat{\bB}_{eo})$ for each $i \in S^C$. If $\tilde{\bB}$
is any solution to (\ref{eq:restricted}), then $\operatorname{Tr}
\bD^T(\tilde{\bB}-\hat{\bB}_{eo}) \leq 0$ for any $\bD \in \partial
J_{eo}(\hat{\bB}_{eo})$, which implies
$(\tilde{\bB}-\hat{\bB}_{eo})_{i,.}=\tilde{\bB}_{i,.}=0$ for all
$i \in S^C$. As $\hat{\bB}_{\text{restricted}}$ is the unique solution to
(\ref{eq:restricted}), this implies $\tilde{\bB}=\hat{\bB}_{eo}$, so
$\hat{\bB}_{eo}$ is the unique solution to (\ref{eq:edgeoutrescaled}).

\subsection*{Proof of Lemma \ref{lemma:restrictedsubgradient}}
Let $\mathcal{D}_i(\hat{\bB})'$ for $i \in S$ be as in the proof of Lemma
\ref{lemma:fullsubgradient} above.
Optimality of $\hat{\bB}$ implies $0 \in \mathcal{D}_i(\hat{\bB})'$ for each $i
\in S$, i.e.\ for some $\bZ \in \mathcal{Z}$,
\[0=-\frac{1}{n}\bX_{.,i}^T(\bX-\bX_S\hat{\bB})+\theta_n \bZ_{i,.}
+\frac{1}{n}\bX_{.,i}^T(0,\ldots,0,\bX_{.,i}-\bX_S\hat{\bB}_{.,i},0,\ldots,0).\]
Combining this condition across $i \in S$ and recalling
$\bX=(\bX_S,\bX_S\bGamma+\bW)=(\bX_S,\bW)+\bX_S\bB^*$,
\[0=-\frac{1}{n}\bX_S^T(\bX-\bX_S\hat{\bB})+\theta_n\bZ+\frac{1}{n}\bDelta
=-\frac{1}{n}\bX_S^T(\bX_S,\bW)-\hat{\bSigma}_{SS}(\bB^*-\hat{\bB})
+\theta_n\bZ+\frac{1}{n}\bDelta.\]
The lemma follows by rearranging and substituting the definition of $\bU$.

\section{Comparison of false detection rates}\label{app:comparison}
\begin{figure}[H]
 \begin{center}
 \includegraphics[width=1.1\textwidth]{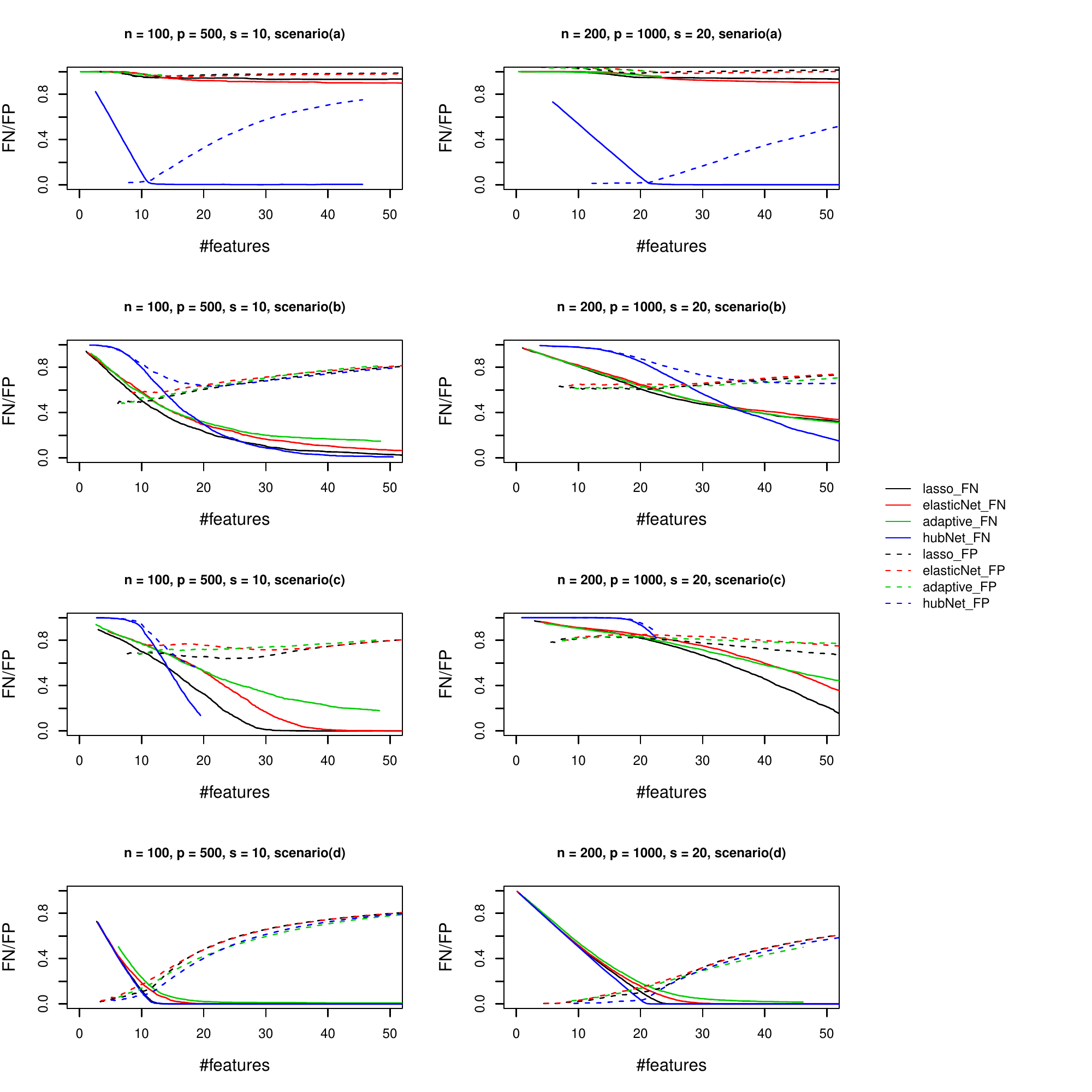}
 \end{center}
 \caption[fig:paths]{\em False positive and false negative  paths under four generating models.}
 \label{fig:paths}
 \end{figure}

\bibliographystyle{agsm}
\bibliography{tibs,extra}

\end{document}